\documentclass{JHEP3}
\usepackage{epsfig}
\usepackage{graphicx}
\usepackage{dcolumn}
\usepackage{bm}
\newcommand{\be}{\begin{equation}}
\newcommand{\ee}{\end{equation}}
\def\bea{\begin{eqnarray}}
\def\eea{\end{eqnarray}}

\usepackage{amssymb}
\usepackage{amsmath}
\usepackage{amsfonts}
\usepackage{graphicx}
\usepackage{psfrag}

 \def\be{\begin{equation}}
\def\ee{\end{equation}}
\def\bea{\begin{eqnarray}}
\def\eea{\end{eqnarray}}
\def\lesssim{\mathrel{\hbox{\rlap{\hbox{\lower4pt\hbox{$\sim$}}}\hbox{$<$}}}}
\def\gtrsim{\mathrel{\hbox{\rlap{\hbox{\lower4pt\hbox{$\sim$}}}\hbox{$>$}}}}

\title{The Non-BPS
Black Hole Attractor Equation }
 \author{ Renata Kallosh$^a$\footnote{\mbox{
Email: {\tt kallosh@stanford.edu} }}, Navin
Sivanandam$^{a,b}$\footnote{\mbox{ Email: {\tt navins@stanford.edu}
}} and Masoud Soroush$^{a,b}$\footnote{\mbox{ Email: {\tt
soroush@stanford.edu} }}
\\
$^a$Department of Physics, Stanford University, Stanford, CA 94305,
USA \\
$^b$SLAC, Stanford University, Stanford, CA 94309, USA}
\received{\today} 
 \preprint{SU-ITP-06/03 \\
  SLAC-PUB-11660\\ \today}
\abstract{ We study the attractor mechanism for extremal non-BPS
black holes with an infinite throat near horizon geometry,
developing, as we do so, a physical argument as to why such a
mechanism does not exist in non-extremal cases. We present a
detailed derivation of the non-supersymmetric attractor equation.
This equation defines the stabilization of moduli near the black
hole horizon: the fixed moduli take values specified by electric and
magnetic charges corresponding to the fluxes in a Calabi Yau
compactification of string theory. They also define the so-called
double-extremal solutions. In some examples, studied previously by
Tripathy and Trivedi, we solve the equation and show that the moduli
are fixed at values which may also be derived from the critical
points of the black hole potential.}
\begin{document}
\section{Introduction}
The supersymmetric black hole attractor mechanism  was introduced in
\cite{Ferrara:1995ih}-\cite{Ferrara:1996dd} and studied in the
context of string theory in \cite{Moore:2004fg} and
\cite{Denef:2001xn}. The possibility of non-supersymmetric black
holes attractors was initially introduced in \cite{Ferrara:1997tw},
where the concept of the effective ``black hole potential'' was
proposed. In general such a potential is a function of electric and
magnetic charges and scalar-dependent vector couplings. For ${\cal
N}=2$ supergravity $V_{BH}= |{\cal D}Z|^2+|Z|^2$, where $Z$ is the
black hole central charge, the charge of the graviphoton.

The attractor mechanism for non-supersymmetric black holes was
recently studied in \cite{Goldstein:2005hq} and
\cite{Tripathy:2005qp}; examples were given and some important
special features were described.  In supersymmetric black hole
attractors the moduli near the black hole horizon are always
attracted to their fixed values since the attractor point is the
minimum of the potential (\cite{Ferrara:1997tw}). For
non-supersymmetric black holes the critical point of the black hole
potential may not be a minimum --- it was stressed in
\cite{Goldstein:2005hq} and \cite{Tripathy:2005qp} that only the
critical points of the potential which are a minima of the black
hole potentials are the true attractor points.

In this paper we shall explore various aspects of non-BPS
attractors. We commence, in Section 2, by analyzing the differences
between extremal (zero temperature) and non-extremal black holes. By
developing an analysis of the scalar field dynamics in both
situations, we are able to construct a physical argument as to why
one would expect only extremal black holes to attract. The essential
geometric difference between the two cases is that extremal black
holes possess an infinite throat where the physical distance to the
horizon is infinite; this is in contrast to the non-extremal case,
where this physical distance is finite. Since the distance acts as
an evolution parameter for the scalar fields, it is only in the
former case that the field can reach its attractor value, and forget
its initial conditions.

In Section 3 we analyze the double-extremal black hole introduced in
\cite{Kallosh:1996tf} and further developed in \cite{Behrndt:1996hu}
and \cite{Behrndt:1996jn}. These solutions have everywhere-constant
scalar fields, a simplification that allows us to study them in some
detail.

It was shown in \cite{Kallosh:2005ax} that in effective ${\cal N}=2$
supergravity, instead of solving the equation for the extremum of
the black hole potential, one can use the attractor equation in the
form:
\begin{equation}
H_3= 2 {\rm Im} \left [  \,  Z \, \overline { \Omega}_3 + \overline
{\mathcal{D}}_{ \bar a}  \overline Z G^{\bar a a}
 {\mathcal{D}}_{ a}  { \Omega}_3 \right ]_{ \partial V_{BH} =0}
\label{nonsusyBH}
\end{equation}
The detailed form of the equation $\partial V_{BH} =0$ was presented
in \cite{Ferrara:1997tw} and it requires that $2(\mathcal{D}_a Z)
\bar Z + i C_{abc}G^{b\bar b} G^{c\bar c} \bar{\mathcal{D}}_{\bar b}
\bar Z \bar{\mathcal{D}}_{\bar c} \bar Z =0$. This form can be used
to make (\ref{nonsusyBH}) explicit. This equation has been already
tested in an example of a non-BPS black hole in
\cite{Giryavets:2005nf}. In Section 4 we use this to establish a
convenient and useful form of the non-BPS black hole attractor
equation:
\begin{eqnarray}\label{attr}
H_3=2\mbox{Im}\left [Z\, \overline { \Omega}_3- \frac{(\overline
{\mathcal{D}}_{\bar a}\overline {\mathcal{D}}_{\bar b} \overline Z)
G^{\bar{a} a}G^{\bar b b }{\mathcal{D}}_{{b}}{Z} }{2{Z}}
{{\mathcal{D}}}_{{a} } { \Omega}_3\right]\ .
\end{eqnarray}

In Section 5 we confirm the validity of the above by solving it
directly for some of the examples from \cite{Tripathy:2005qp}. In
these examples the attractor points were identified in
\cite{Tripathy:2005qp} by solving the equation $\partial_a V_{BH}
=0$. Here we will solve equation (\ref{attr}) and find both
supersymmetric attractors with ${\mathcal{D}}_{{c}}{Z}=0$ and the
non-supersymmetric ones with ${\mathcal{D}}_{{c}}{Z}\neq 0$. In the
supersymmetric case the second term in  (\ref{attr}) vanishes.
However, in the non-supersymmetric case it conspires with the first
term so that the moduli have to become function of charges to
satisfy the equation. We also present an example of a double
extremal non-BPS black hole.

Finally, we discuss the relation between non-BPS attractors and the
O'Raifeartaigh model of spontaneous SUSY-breaking. In both cases the
system cannot, by design, go to the supersymmetric minimum and
therefore is stable.

\section{Extremality and Attractiveness}
Here we present the main features of the stabilization of moduli
near a black hole horizon. We confirm, with a detailed explanation,
that such stabilization is not necessarily related to unbroken
supersymmetry and, in fact, that the existence of attractor-like
behavior depends on the extremality (or otherwise) of the black
hole.

We follow \cite{Ferrara:1997tw} and start by writing down the
bosonic part of the Einstein-Maxwell action coupled to some Abelian
vector fields:
\begin{equation}
-{R\over 2} + G_{a\bar a} \partial_ \mu z^a   \partial_\nu \bar
z^{\bar a} g^{\mu\nu} +
 {\rm Im} {\cal N}_{\Lambda \Sigma} {\cal F}^{\Lambda}_{\mu \nu}  {\cal
F}^{ \Sigma}_{\lambda \rho}  g^{\mu \lambda} g^{\nu \rho} + {\rm Re}
{\cal N}_{\Lambda \Sigma} {\cal F}^{\Lambda}_{\mu
\nu}\left(\ast{\cal F}^{ \Sigma}_{\lambda \rho}\right) g^{\mu
\lambda} g^{\nu \rho}
 \ .
\label{scalaraction2}
\end{equation}
This action may have an arbitrary scalar metric $G_{a\bar a}$ and
arbitrary scalar dependent vector couplings  ${\rm Re} {\cal
N}_{\Lambda \Sigma}$ and  ${\rm Im} {\cal N}_{\Lambda \Sigma}$. In
the special case that the bosonic action is part of ${\cal N}=2$
supergravity action, the positive definite metric ($G_{a\bar a}$) on
the scalar manifold and the scalar dependent negative definite
vector couplings (${\rm Re} {\cal N}_{\Lambda \Sigma}$ and  ${\rm
Im} {\cal N}_{\Lambda \Sigma}$) can be extracted from the
prepotential or the symplectic section that defines a particular
${\cal N}=2$ theory. Such an effective action can also be derived
from the compactification of string theory on Calabi-Yau manifolds.

We consider the following static, spherically symmetric ansatz  for
the metric
\begin{equation}
ds^2 = e^{2U} dt^2 - e^{-2U} \left [{c^4  \over \sinh^4 c\tau} d
\tau^2 + {c^2 \over \sinh^2 c\tau}d \Omega^2 \right]\ . \label
{ansatz}
\end{equation}
It was shown in \cite{Ferrara:1997tw}  that the effective
1-dimensional Lagrangian from which the radial equations for
$U(\tau), z(\tau)$ and $\bar z(\tau)$ as well as electric
($\psi^\Lambda(\tau)$) and magnetic ($\chi_\Lambda(\tau)$)
potentials may be derived, is a pure geodesic action
\begin{equation}
\hat G_{ij} {d \hat \phi ^i \over d\tau } {d \hat \phi ^j  \over
d\tau }\ ,
\end{equation}
with the constraint
\begin{equation}
\hat G_{ij} {d \hat \phi ^i \over d\tau } {d \hat \phi ^j  \over
d\tau } = c^2.
\end{equation}
Here the hatted fields include $U(\tau), z(\tau), \bar z(\tau),
\psi^\Lambda(\tau), \chi_\Lambda(\tau)$. Taking into account the
gauge invariance of the vector multiplet part of the action one can
express the derivatives of the electric and magnetic potentials via
conserved electric and magnetic charges. The resulting
one-dimensional Lagrangian for the evolution of $U(\tau), z(\tau)$
and $\bar z(\tau)$ is not pure geodesic anymore, it now has a
``black hole potential'':
\begin{equation}
 {\cal L} \left (U(\tau) , z^a(\tau) , \bar z^{\bar a}(\tau) \right )= \left ({d U
\over d\tau}\right )^2 +  G_{a\bar a} {dz^a \over d \tau} {d \bar
z^{\bar a} \over d \tau}
 + e^{2U} V_{BH}(z,\bar z, p,q)
\ . \label{lagr}
\end{equation}
The constraint, in turn, becomes:
\begin{equation}
  \left ({d U
\over d\tau}\right )^2 +    G_{a\bar a} {dz^a \over d \tau} {d \bar
z^{\bar a} \over d \tau}
 - e^{2U} V_{BH}(z,\bar z, p,q) =c^2
\ . \label{constr1}
\end{equation}
Further details of this calculation can be found in Appendix A. In
general $V_{BH}$ is an expression that depends on the charges and
the vector couplings; its explicit form is given in equations
(12)and (13) of \cite{Ferrara:1997tw}. In case of ${\cal N}=2$
supergravity we have:
\begin{equation}
V_{BH}(z,\bar z, p,q)=\left( |Z(z,\bar z, p,q)|^2 + |{\cal D}_a
Z(z,\bar z, p,q)|^2 \right)\ . \label{pot1}
\end{equation}
$Z$ is the central charge, the charge of the graviphoton in ${\cal
N}=2$ supergravity and ${\cal D}_a Z$ is the K\"{a}hler covariant
derivative of the central charge (some details of our notation for
derivatives are discussed in Appendix B):
\begin{equation}
Z(z, \bar z, q,p) = e^{K(z, \bar z)\over 2} (X^\Lambda(z)  q_\Lambda
- F_\Lambda(z) \, p^\Lambda)= (L^\Lambda q_\Lambda - M_\Lambda
p^\Lambda) \ . \label{central}
\end{equation}
Here  $c^2 = 2ST$, where $S$ is the entropy and $T$ is the
temperature of the black hole. At infinity, as $\tau \rightarrow 0$,
$U\rightarrow M\tau$ and one finds a Minkowski metric and the
constraint:
\begin{equation}
M ^2 (z_{\infty}, \bar z_{\infty} , p,q ) -   |Z(z_{\infty} ,\bar
z_{\infty}, p,q)|^2  =c^2+ |{\cal D}_a Z(z_{\infty},\bar z_{\infty},
p,q)|^2- G_{a\bar a } \Sigma^a \overline {\Sigma}^ {\bar a} \ .
\end{equation}
The dilaton charge at infinity is defined as $\Sigma^a=\left ( {d
z^{a} \over d \tau}\right )_{\infty}$. The BPS configuration has its
mass equal to the central charge in supersymmetric theories so that:
\begin{equation}
M ^2 (z_{\infty}, \bar z_{\infty} , p,q ) =    |Z(z_{\infty} ,\bar
z_{\infty}, p,q)|^2   \ ,   \qquad  c=0  \ ,   \qquad G^{a \bar a}
\overline {\cal D}_{\bar a}  Z(z_{\infty} ,\bar z_{\infty}, p,q)=
\Sigma^a \ .\label{BPS}
\end{equation}
In this paper, following \cite{Goldstein:2005hq} and
\cite{Tripathy:2005qp}, we are primarily interested in non-BPS
solutions where the 1st order BPS equation $\left ( {d z^{a} \over d
\tau}\right )= G^{a \bar a} \overline {\cal D}_{\bar a}  Z(z ,\bar
z, p,q)$ is not satisfied.

These solutions can be divided into two classes:
\begin{enumerate}
\item The first case, with $c^2= 2ST\neq 0$ in  (\ref{ansatz}) and
${\cal D}_a Z(z ,\bar z, p,q)\neq 0$, describes non-BPS black holes
with surface gravity $>0$.  When these are charged they are
non-extremal, they have two non-coincident horizons  and a
non-vanishing temperature. They can evaporate quantum-mechanically
until they reach zero temperature and consequent extremality.
\item The second type is the end stage of the evaporation described above. Here $c^2= 2ST = 0$, but
the black holes still have ${\cal D}_a Z(z ,\bar z, p,q)\neq 0$.
Thus they are extremal and at zero temperature, but they are still
not-BPS and have no unbroken supersymmetry.
\end{enumerate}

Let's proceed by considering the latter type of black hole. Although
$c=0$, in contrast to (\ref{BPS}) the 1st order BPS equation $\left
( {d z^{a} \over d \tau}\right )= G^{a \bar a} \overline {\cal
D}_{\bar a} Z(z ,\bar z, p,q)$ is not satisfied for these solutions.
Therefore $M^2\neq |Z|^2$, and instead:
\begin{equation}
M ^2 (z_{\infty}, \bar z_{\infty} , p,q )
 -   |Z(z_{\infty} ,\bar z_{\infty}, p,q)|^2  = |{\cal D}_a Z(z,\bar z, p,q)|^2-  |\Sigma^a |^2 > 0
\ . \label{nonBPS}
\end{equation}
We can also use (\ref{ansatz}) to obtain an expression for the
geometry at $c=0$:
\begin{equation}
ds^2 =  e^{2U} dt^2 - e^{-2U} \left[{d\tau^2 \over \tau^4} + {1
\over \tau^2} (d \theta ^2 +  \sin^2 \theta d \varphi ^2)\right]  \
.
\end{equation}
Requiring that the solution has finite horizon area leads us to
conclude:
\begin{equation}
e^{-2U}\rightarrow \left( {A\over 4\pi}\right) \tau^2  \qquad {\rm
as} \quad \tau  \rightarrow - \infty \ .
\end{equation}
Thus the near horizon geometry is given by:
\begin{equation}
ds^2 =   {4\pi \over A\tau^2}  dt^2 -   \left( {A\over 4\pi}\right)
\left[{d\tau^2 \over \tau^2} + (d \theta ^2 +  \sin^2 \theta d
\varphi ^2)\right]  \ .
\end{equation}
Using $\rho= -{1\over \tau}$ and $\omega = \log \rho$ this becomes
the Bertotti-Robinson product space, $AdS_2\times S^2$:
\begin{equation}
ds^2 = \left(  {4\pi \over A} \right ) e^{2\omega} dt^2 -   \left(
{A\over 4\pi}\right) d\omega^2 -   \left(  {A\over 4\pi}\right) d
\Omega^2  \ . \label{bertotti}
\end{equation}
We can repeat this analysis for non-extremal ($c\neq0$), non-BPS
black holes. To start we consider the limit of the geometry at the
horizon as $\tau \rightarrow -\infty$ and generalize our earlier
observation about the need for a finite area solution:
\begin{equation}
e^{-2U}\rightarrow  {A\over 4\pi} {\sinh^2 c\tau \over c^2}  \qquad
{\rm as} \quad \tau  \rightarrow - \infty \ .
\end{equation}
The near horizon geometry (for arbitrary $c$) then becomes:
\begin{equation}
ds^2 =  {4\pi c^2\over A \sinh^2 c\tau}  dt^2 - {A\over 4\pi}  {c^2
\over \sinh^2 c\tau}d\tau^2 - {A\over 4\pi}d^2\Omega  \ .
\end{equation}
The above can be represented as:
\begin{equation}
ds^2 =     \left ({4 \pi \over A}\right) c ^2 \rho^2 dt^2 - \left
({A \over 4 \pi}\right) d\rho^2 - \left ({A \over 4 \pi}\right) d
\Omega^2   \ , \qquad  \rho  \rightarrow 0 \ . \label{c}
\end{equation}
We have used the approximation that $\sinh c\tau \rightarrow
-e^{-c\tau}/2$ as $\tau  \rightarrow - \infty $ and also changed
variables to $\rho= 2 e^{c\tau}$. We can get the same near horizon
geometry without making this approximation, but instead performing a
change of variables $x= \log (-\tanh {c\tau\over 2})$. This will
lead to the metric:
\begin{equation}
ds^2 =  \left ({4 \pi \over A}\right) c^2 \sinh x^2 dt^2 - \left ({A
\over 4 \pi}\right)  dx^2 - \left ({A \over 4 \pi}\right)  d
\Omega^2   \ , \qquad  x  \rightarrow 0 \ . \label{x}
\end{equation}
If, as we approach the horizon, we take the limit $x\rightarrow 0$,
we will reproduce the geometry in  (\ref{c}). With a re-scaling of
$t$ and defining $r_h^2=\left(\frac{A}{4\pi}\right)$ we can write
the above as:
\begin{equation}
ds^2 =  \rho^2dt^2 - (r_h)^2 \,  d \rho^2 - (r_h)^2 \, d \Omega^2 \
, \qquad \rho\rightarrow 0 \ . \label{nonextremal}
\end{equation}
Note that the coordinate $\rho$ is the {\it physical} distance to
the horizon in the units of $r_h$ -- that is to say at any given
time the interval $ds^2$ is equal to coordinate distance
$(r_h)^2d\rho^2$. To put this another way; if one starts at some
finite values $\rho_0$ the physical distance to the horizon is given
by:
\begin{equation}
\Delta \rho= \rho_0 - \rho_h = \rho_0 \ . \label{finite}
\end{equation}
This is finite.  Now let's compare this with the near horizon
geometry for extremal black holes with the $AdS_2\times S^2$
geometry:
\begin{equation}
ds^2 =   { e^{2\omega}\over (r_h)^2 }\,  dt^2 -   (r_h)^2 \,
d\omega^2 -  (r_h)^2 \,  d \Omega^2  \ , \qquad \omega_h \rightarrow
-\infty\ . \label{extremal}
\end{equation}
This expression follows from (\ref{bertotti}), after making the
above replacement for $r_h$. In this metric if one starts at some
finite value of the {\it physical} distance coordinate $\omega_0 $,
one finds that the distance to the horizon is infinite.
\begin{equation}
\Delta \omega = \omega_0 - \omega_h   \rightarrow \infty
\end{equation}
This physical difference in the near horizon geometries can give us
some considerable insight into why the latter set of black holes can
have attractors, while the former cannot. The infinite distance to
the horizon in the extremal black hole case is, of course,
characteristic of so-called infinite throat geometries. To see how
this leads to the attractor mechanism it is helpful to consider a
classical mechanical system. In such a system attractor behavior
follows when a fixed point ($x_{fix}$ such that $v(x_{fix})=0$) of a
motion $x(t)$ is reached in the limit $t\rightarrow \infty$. In our
gravitational example the role of the evolution parameter $t$ is
played by physical distance to the horizon. Proceeding a little
further with this line of reasoning it is clear that the infinite
physical distance is key to allowing a scalar field to forget about
its initial conditions\footnote{We are grateful to A. Linde who
suggested this explanation of attractor/non-attractor behavior.}.
For whilst in the non-extremal case the field only has finite
``time'' until it reaches the horizon (ensuring the lingering memory
of initial conditions), the presence of an infinite throat provides
the guarantee that the scalar will be captured by the inexorable
lure of the attractor\footnote{A somewhat incomplete, but perhaps
helpful, analogy would be an under-damped oscillator. In any finite
time interval the position and velocity will be related to the
initial conditions. However, as $t\rightarrow\infty$ the oscillator
settles at its equilibrium point.}.

\begin{figure}[h!]
\centering{\includegraphics[width=0.48\textwidth,height=0.3\textheight]{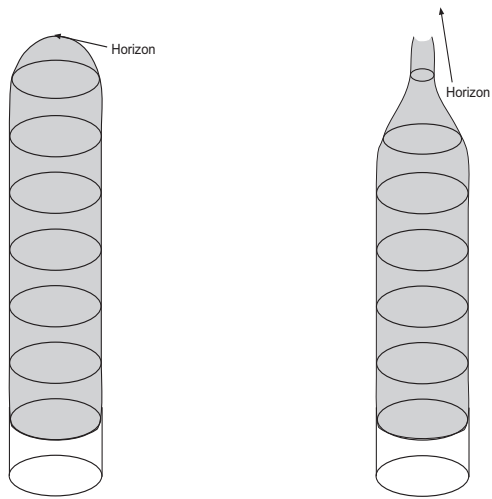}}
\caption{On the left there is an Euclidean section of the near
horizon ``cigar'' geometry of the non-extremal black holes with
non-vanishing temperature. On the right there is an Euclidean
section of an infinite throat near horizon geometry of the  extremal
black holes with vanishing temperature. }\label{fig:Fig1}
\end{figure}

Euclidean spatial slices of this geometry are illustrated in Figure
\ref{fig:Fig1} (analogous to those in \cite{Gibbons:1994ff}. These
clearly show the distinct differences  in the near horizon
geometries of our examples.

To transform this argument into something more rigorous, we will
basically repeat the reasoning given in \cite{Ferrara:1997tw} that
if extremal black holes have a non-singular geometry and regular
values of scalars at the horizon, then such scalars will take the
universal values defined by the minimum of the black hole potential.
We will use the near horizon geometry with the physical distance
coordinate $\omega$ which at the horizon goes to $-\infty$; and, as
expected, we will show that source of this universality is
attributed to the infinite physical distance to the horizon. An
analogous derivation in the non-extremal black holes near horizon
geometry  with the physical distance coordinate $\rho$, which at the
horizon goes to $0$, will show why the horizon values of scalars are
non-universal; again, as our reasoning in the previous paragraph
suggests, the finite distance to the horizon will prove to be the
key\footnote{In \cite{Goldstein:2005hq} an argument was given as to
why extremal black holes are attractive and non-extremal ones are
not; examples of scalar field behavior in the latter system (from
\cite{Kallosh:1992ii}) were also plotted. Our approach to this
problem is based on \cite{Ferrara:1997tw} where the authors work
with coordinate systems that use the physical distance. This allows
us to see the difference between these two systems as the result of
infinite vs finite physical distance to the horizon.}.

We start by deriving the equations of motion for the scalar field
from the Lagrangian:
\begin{equation}
 {\cal L} \left (U(\tau) , z^a(\tau) , \bar z^{\bar a}(\tau) \right )= \left(\left ({d U
\over d\tau}\right )^2 +  G_{a\bar a} {dz^a \over d \tau} {d \bar
z^{\bar a} \over d \tau}
 + e^{2U} V_{BH}(z,\bar z, p,q)\right)
\ . \label{lagr1}
\end{equation}
Varying this action with respect to $\bar{z}^{\bar{a}}$) gives:
\begin{equation}
G_{a\bar{b}}\partial_{\tau}\left(\partial_{\tau}z^a\right)+\left(\frac{\partial
G_{a\bar{b}}}{\partial \bar{z}^{\bar{a}}}-\frac{\partial
G_{a\bar{a}}}{\partial
\bar{z}^{\bar{b}}}\right)\partial_{\tau}\bar{z}^{\bar{a}}\partial_{\tau}z^a+\frac{\partial
G_{a\bar{b}}}{\partial
z^b}\partial_{\tau}z^a\partial_{\tau}z^b=e^{2U}\frac{\partial
V}{\partial\bar{z}^{\bar{b}}}\ .
\end{equation}
If we assume that that the moduli space is a complex K\"{a}hler
manifold with a K\"{a}hler metric $G_{a\bar{b}}$, this simplifies
to:
\begin{equation}
\partial_{\tau}\left(\partial_{\tau}z^a\right)+\Gamma^a_{bc}(z,\bar{z})\partial_{\tau}z^b\partial_{\tau}z^c=
G^{a\bar{b}}e^{2U}\frac{\partial V}{\partial\bar{z}^{\bar{b}}}\ .
\label{eqofmtn}
\end{equation}
Note the the assumption of K\"{a}hlerity is not essential and the
following arguments hold in general. However, for reasons of
clarity, we will work in this more constrained situation.

\noindent {\it Extremal black holes}

\noindent The equation of motion (\ref{eqofmtn}) for the scalars in
the near horizon geometry (\ref{extremal}) is given by:
\begin{equation}
 \partial_{\tau}\left( \partial_{\tau}z^a\right) +
\Gamma^a_{bc} (z,\bar{z})\partial_{\tau}z^b
\partial_{\tau}z^c= G^{a\bar{b}} e^{2U}\frac{\partial V}{\partial\bar{z}^{\bar{b}}}\ . \label{scalareq}
\end{equation}
We have set $r_h=1$ for simplicity. Using
$\partial_{\tau}=(-1/\tau)\partial_{\omega}$ and
$e^{2U}\rightarrow(r_h^2/\tau^2)$ we obtain:
\begin{equation}
(z^{a})''+(z^a)'+\Gamma^a_{bc}(z^b)'(z^c)'=G^{a\bar{b}}\frac{\partial
V}{\partial \bar{z}^{\bar{b}}}\ , \label{extremalcondition}
\end{equation}
where $(z^a)'\equiv\partial_{\omega}z^a$.

We now argue that the l.h.s. of this expression must vanish
identically at the horizon. To see this recall that we are working
with the \textit{physical} distance as our coordinate. Accordingly
we expect that the scalar field and all its derivatives with respect
to this coordinate will be finite and tend to a definite limit at
the horizon. However, if, say, the first derivative of $z^a$ tends
to some non-zero limit as $\omega\rightarrow-\infty$ then
$z^a\rightarrow\infty$ as we approach the horizon. Hence, if
$z^a=constant$ at the horizon, then all the derivatives must be zero
there (this extends to higher derivatives by induction).

We are left, then, with the following condition at the horizon:
\begin{equation}
\left.\frac{\partial V}{\partial z^a}\right|_{z^a_h}=0\ .
\end{equation}
Since the values of the scalar field that solve this equation are
independent of their initial conditions, the horizon is an attractor
where the scalar field takes values $z^a_h\equiv z^a_h(p,q)$.

Before moving on let's take a brief detor. It is clear that the
analysis above will still hold if the black hole in question is BPS.
Of course in this situation we will have the following additional
conditions:
\begin{eqnarray}
\label{M=Z}{d U \over d\tau}&=&e^U |Z|\ , \\
\label{DZ=0} {d z^a  \over d\tau } &=& e^{U} G^{a\bar a} \bar {\cal
D}_{\bar a} \bar Z\ .
\end{eqnarray}
These follow immediately from (\ref{constr1}), (\ref{pot1}) and the
1st order BPS condition. As $\tau\rightarrow-\infty$  (\ref{DZ=0})
can be given in the form (with $r_h=1$):
\begin{equation}
{d z^a \over d\omega } = G^{a\bar a} \bar {\cal D}_{\bar a} \bar Z\
.
\end{equation}
Therefore, at the horizon, where the moduli are stabilized, we have:
\begin{equation}
{d z^a  \over d\omega } = G^{a\bar a} \bar {\cal D}_{\bar a} \bar
Z=0\ ,
\end{equation}
and the number of unbroken supersymmetries is doubled.
\\

\noindent {\it Non-extremal black holes}

\noindent Let's now examine the equation of motion (\ref{eqofmtn})
in the non-extremal near horizon geometry (\ref{nonextremal}):
\begin{equation}
\partial_{\tau} \left(\partial_{\tau}z^a\right)
+ \Gamma^a_{bc}(z,\bar{z}) \partial_{\tau}z^b
\partial_{\tau}z^c= G^{a\bar{b}} e^{2U}\frac{\partial V}{\partial\bar{z}^{\bar{b}}}\ . \label{scalareq2}
\end{equation}
Making the substitutions $e^{2U}\rightarrow(-c^2\rho^2/r_h^2)$ and
$\rho=2e^{c\tau}$ we obtain:
\begin{equation}
\rho(z^a)''+(z^a)'+\rho\Gamma^a_{bc}(z^b)'(z^c)'=-\rho
G^{a\bar{b}}\frac{\partial V}{\partial\bar{z}^{\bar{b}}}\ .
\label{scalarnon extremal}
\end{equation}
The prime denotes a derivative with respect to $\rho$. This in turn
implies that at $\rho=0$:
\begin{equation}
(z^a)'=0\ .
\end{equation}
Substituting this gives us the following equation at the horizon:
\begin{equation}
\left.(z^a)''\right|_{\rho=0}=\left.-G^{a\bar{b}}\frac{\partial
V}{\partial\bar{z}^{\bar{b}}}\right|_{z^a_h}\ .
\end{equation}
While at first glance this may seem similar to the equation obtained
for the extremal case (\ref{extremalcondition}), there is an
important difference. Here the coordinate $\rho$ takes the value 0
in the horizon limit. Thus, if we insist that the field and all its
derivatives with respect to $\rho$ are finite at the horizon (since
it is a physical coordinate), this does not place any further
constraints on the $z^a$. Indeed such an assertion simply means that
the scalar field has a Taylor expansion around $\rho=0$. Thus:
\begin{equation}
\left.-G^{a\bar{b}}\frac{\partial
V}{\partial\bar{z}^{\bar{b}}}\right|_{z^a_h}=A_2^a\ .
\end{equation}
Here $A_2^a$ are the appropriate coefficients of the Taylor
expansion and are, generically, dependent on some initial conditions
and not equal to zero. Therefore, for non-extremal black holes the
values that the scalar fields take will depend on their initial
conditions --- there is no attractor.

\section{Double-extremal Black Holes}

\begin{figure}[h!]
\centering{\includegraphics[width=0.55\textwidth,height=0.3\textheight]{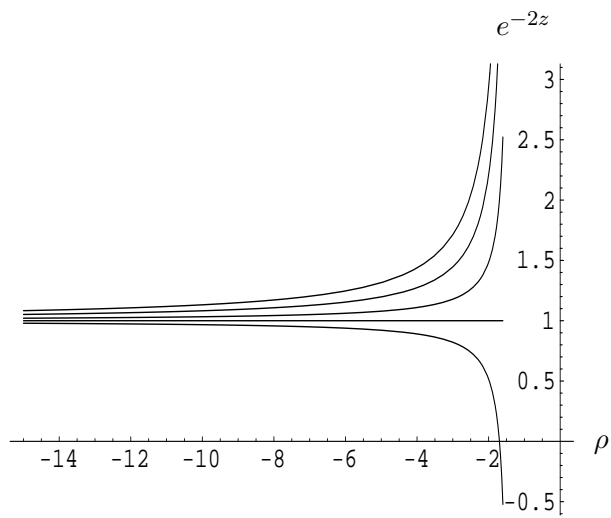}}
\caption{Evolution of the dilaton field $e^{-2z}$ for various
initial conditions at $\rho = 0$ ($r = \infty$) to a common fixed
point at $\rho =-\infty$ ($r = 0)$ where $\rho= -{1\over r}$. The
special case of the horizontal line represents a double-extremal
black hole.} \label{F22}
\end{figure}

Double-extremal BPS black holes  have everywhere-constant moduli,
they were studied in \cite{Kallosh:1996tf}, \cite{Behrndt:1996hu}
and \cite{Behrndt:1996jn}. For example consider Figure \ref{F22}
where we have plotted the evolution of the dilaton as a function of
the coordinate $\rho$ for a particular extremal black hole example
from \cite{Kallosh:1992ii}. This coordinate $\rho= -{1\over r}$ is
inverse to isotropic coordinate $r$. The double-extremal case
corresponds to fixed values of the moduli everywhere in space-time
--- it is represented by the horizontal line in the plot.

The black hole solution presented in \cite{Kallosh:1996tf} may be
supersymmetric if the constant scalars satisfy the equation ${\cal
D}Z=0$. However, the solution is valid also when this condition on
scalars is not satisfied,  as long as $\partial_a V=0$. The theory
defined by the action (\ref{scalaraction2}) has a solution with the
extremal Reissner-Nordstr\"{o}m geometry and fixed scalars whose
values are defined by the condition  $\partial_a V=0$. As we will
show in the next section, this condition is also a consequence of
the general attractor equation, and is valid for supersymmetric as
well as for non-supersymmetric solutions.

A point that should be stressed here is that in the
Reissner-Nordstr\"{o}m case there is only one vector field, a
graviphoton,  and that the vector coupling ${\cal N}_{\Lambda
\Sigma}$ in
 (\ref{scalaraction2}) is trivial, with only $\rm \, Im
{\cal N}_{00}= -1/2$  for 00 component --- there are no axions. In
fact, even more general cases double-extremal black holes have a
similar metric. There are differences though; the constant scalars
now include axions, the matrix ${\cal N}_{\Lambda \Sigma}$ depends
on $(p,q)$ and the full solution has a multi-component vector field.

\psfrag{eminus2phi}{$e^{-2z}$} \psfrag{rho}{$\rho$}

We call these black holes double-extremal since they pick up the
values of the moduli at infinity that extremize the black hole mass;
values that are equal to those at the horizon. There is no energy
stored in the scalars, so ${\partial z\over \partial \tau}=0$
everywhere. While for the BPS case  ${\cal D}Z=0$ everywhere, the
most general conditions are that double-extremal black holes (BPS or
otherwise) have Reissner-Nordstr\"{o}m geometry and everywhere
constant scalars whose values are defined by the critical point of
the black hole potential:
\begin{equation}
\partial_a V_{BH}=0 \ , \qquad z_{fix}(p,q) = z_{\infty}= z_{h}\ .\label{debhcond}
\end{equation}
Obviously for BPS states (\ref{debhcond}) is automatically
satisified, since $\mathcal{D}_aZ=0\Rightarrow\partial_a V_{BH}=0$
The double-extremal BPS and non-BPS black hole solution is given in
isotropic coordinates by the metric:
\begin{equation}
ds^2 = e^{2U} dt^2 - e^{-2U} d\vec x^2\ .
\end{equation}
Where:
\begin{equation}
 e^{-U}(r) = 1+{\sqrt{ A/4\pi}\over r}= 1+{M\over r}\ .
\label{dextrmetric}
\end{equation}
With $r^2=\vec x^2$ and the horizon at $r=0$. The constant $A$ is
the area of the horizon defined by the value of the potential at the
horizon (and, for that matter, the value everywhere else):
\begin{equation}
{ A\over 4\pi}  =   V(z_h, \bar z_h, p,q)= V(z_{\infty} ,\bar
z_{\infty}, p,q))=V(z_{fix}, \bar z_{fix}, p,q)\ .
\end{equation}
The vanishing scalar charges, $\Sigma^a$, imply that the the
constraint equation at infinity takes the form:
\begin{equation}
{ A\over 4\pi}= M ^2 (z_{\infty}, \bar z_{\infty} , p,q ) = V
(z_{fix} ,\bar z_{fix}, p,q) \ .
\end{equation}
The vector field can be defined by the symplectic doublet $( {\cal
F}^\Lambda, \, {\cal G}_\Lambda) $. Here ${\cal G}_\Lambda$ is
related to  the field ${\cal F}^\Lambda$ and its dual  $*{\cal
F}^\Lambda$ as follows: ${\mathcal{G}}_\Lambda= (\rm Re \, {\cal N}
\, {\cal F})_\Lambda - (\rm Im \, {\cal N} \,(* {\cal F}))_\Lambda$.
The double-extremal solution for the  vector fields  is given by
\cite{Kallosh:1996tf} as:
\begin{equation}
{\cal F}^\Lambda= e^{2U} {2Q^\Lambda\over r^2} dt\wedge dr-
2P^\Lambda d\theta \wedge \sin \theta d\phi\ . \label{vector}
\end{equation}
The ``dressed'' electric and magnetic charges $Q^\Lambda, P^\Lambda$
which are involved in (\ref{vector}) are related to quantized
electric and magnetic charges $q_\Lambda, p^\Lambda$ via the moduli
dependent vector couplings ${\cal N}_{\Lambda \Sigma}$ as follows:
 \begin{eqnarray}\label{dressed}
\left(\begin{array}{cc} P^\Lambda \\
Q^\Lambda \end{array}\right)= {1\over 2} \left(\begin{array}{cc} p^\Lambda \\
((\rm Im \, {\cal N})^{-1} (\rm Re \, {\cal N}) p)^\Lambda- ((\rm Im
\, {\cal N})^{-1}q)^\Lambda \end{array}\right)\ .
\end{eqnarray}
The quantized charges $(p^\Lambda,q_\Lambda)$ can be identified with
magnetic charges of the doublet $( {\cal F}^\Lambda, \, {\cal
G}_\Lambda) $ via the surface integral:
\begin{equation}
\left(\begin{array}{cc} p^\Lambda \\
q_\Lambda \end{array}\right)=  \left(\begin{array}{cc}\int  {\cal F}^\Lambda \\
\int  {\cal G}_\Lambda\end{array}\right)\ .
\end{equation}
Further, we can find the vector couplings as functions of the moduli
coordinates. In $\mathcal{N}=2$ supergravities defined by the
holomorphic prepotential $F(X)$, the couplings are given in
\cite{deWit:1984px} as:
\begin{eqnarray}\label{vectorcoupling}
{\mathcal{N}}_{\Lambda\Sigma}=\overline{F}_{\Lambda\Sigma}+2i\frac{(\mbox{Im}F_{\Lambda\Omega})
(\mbox{Im}F_{\Pi\Sigma})z^{\Omega}z^{\Pi}}{(\mbox{Im}F_{\Gamma\Delta})z^{\Gamma}z^{\Delta}}\
.
\end{eqnarray}
$F_{\Lambda\Sigma}$ is the second derivative of the prepotential,
$\partial_{\Lambda}\partial_{\Sigma} F$ and the special coordinates
$z^{\Lambda}$ are defined by
$z^{\Lambda}=\frac{X^{\Lambda}}{X^{0}}$, so that
$z^{\Lambda}=(1,{X^{a}\over X^0})$. For a Calabi-Yau moduli space
with $F= D_{abc} {X^a X^b X^c\over X^0}$ one finds
(\cite{Behrndt:1996hu}):
\begin{eqnarray}\label{F-lam-sig}
F_{\Lambda\Sigma}=\left(
\begin{array}{cc}
2D_{abc}z^{a}z^{b}z^{c} & -3D_{abc}z^{b}z^{c} \\
-3D_{abc}z^{b}z^{c} & 6D_{abc}z^{c} \\
\end{array}
\right)\ .
\end{eqnarray}

In models where the prepotential $F$ does not exist and only the
section $(X^\Lambda, F_\Lambda)$ is available, the form of the
vector couplings ${\mathcal{N}}_{\Lambda\Sigma}$ can still be found
--- see \cite{Ceresole:1995jg} for details. In order to completely specify the double-extreme black hole solution at all
values of $r$, one also has to give, in addition to the metric and
the attractor values of the scalars, the values of the vector
fields. We should therefore calculate the vector couplings ${\cal
N}_{\Lambda \Sigma}$ at the attractor point for the scalars,
providing the values of the ``dressed'' electric and magnetic
charges $ Q^\Lambda$ and $P^\Lambda$. In terms of these charges the
fixed value of the black hole potential has a simple expression:
\begin{equation}
V_{fix}(p,q) =  (|Z|^2+|{\cal D}Z|^2)_{fix}= \left[-2 \rm Im \,
{\cal N}_{\Lambda \Sigma}(Q^\Lambda Q^\Sigma + P^\Lambda P^\Sigma
)\right]_{fix} \label{Z2}
\end{equation}
As we established earlier, since double-extremal black holes have
$\partial_a V_{BH}=0$ at all points, including for moduli at
infinity, the sobriquet is valid for BPS and non-BPS black holes.
However, there is a subtle difference --- while for the BPS black
holes the mass is always at a minimum at infinity, for the non-BPS
case one should require (as proposed in \cite{Goldstein:2005hq},
\cite{Tripathy:2005qp}) that the second derivative is positive
definite .

We will give examples of a BPS and a non-BPS double-extremal black
hole in Section 5.2. These examples will be consistent with the
those of the solution of the attractor equation in Section 5.1.

\section{The Attractor Equation}

In this section, we introduce the explicit form of the black hole
attractor equation. The equation we find is valid for
non-supersymmetric cases as well as for supersymmetric ones. The
attractor mechanism in the latter scenario is well known and was
developed in \cite{Ferrara:1995ih}-\cite{Ferrara:1996dd}.

In the context of type-IIB string theory the electric and magnetic
charges of a black hole originate from the self-dual 5-form. Three
legs of this 5-form belong to a Calabi-Yau manifold and the other
two are in the $4d$ space-time.  When the 5-form is integrated over
a supersymmetric 3-cycle in the Calabi-Yau manifold it gives a
field-strength for the effective ${\mathcal{N}}=2$ supergravity in
the $4d$ space-time. The field-strength in $4d$ space-time can, in
turn, be integrated over some appropriate 2-cycle to define the
electric and magnetic charges. If the 5-form is integrated over the
relevant 2-cycle first, we will have a 3-form with regard to the
Calabi-Yau.

To set up a normalization in agreement with \cite{Ferrara:1996dd}
and \cite{Ferrara:1990dp}, we remind ourselves that the graviphoton
field strength, $T_{\mu\nu}^-$, and the vector multiplet field
strength, ${\cal F}^{i-}_{\mu\nu}$,  appear in the $4d$ gravitino
and chiral gaugino transformations as follows:
\begin{eqnarray}
\delta \psi_{A\mu} &=& {\cal D}_\mu \epsilon_A +\epsilon_{AB} T_{\mu\nu}^- \gamma^\nu \epsilon^B \ , \nonumber\\
\delta \lambda^{aA} &=& i\gamma^\mu \partial_\mu z^a \epsilon^A+
{i\over2} {\cal
F}^{a-}_{\mu\nu}\gamma^{\mu\nu}\epsilon_B\epsilon^{AB}\ .
\label{ricca}
\end{eqnarray}
$\epsilon^{A}$ is the parameter of the transformation and
$\epsilon^{AB}$ is the complete antisymmetric tensor. The central
charge and its covariant derivative in ${\mathcal{N}}=2,\ d=4$
supergravity are defined as follows:
\begin{eqnarray}\label{tdod}
&&Z = -{1\over 2} \int_{S_2} T^- \ , \qquad Z_a \equiv {\cal D}_a Z
=
-{1\over 2} \int_{S_2} {\cal F}^{+\bar a} G_{a\bar a}\ ,\\
&&\overline Z = -{1\over 2} \int_{S_2} T^+ \ , \qquad \overline
Z_{\bar a} \equiv \overline {\cal D}_{\bar a} \overline  Z =
-{1\over 2} \int_{S_2} {\cal F}^{- a} G_{a\bar a}\ . \label{tdod}
\end{eqnarray}

In $10d$ supergravity, we start with the self-dual five-form
$\mathcal{F}=*{\cal F}$ on the manifold $M_4\times K_6$. A real
conserved field strength is given by the imaginary part of the
five-form ${\cal F}^+$ such that ${\cal F}_5 = i({\cal F}^- - {\cal
F}^+ )= 2 {\rm Im}\, {\cal F}^+$. The appropriate conservation
equation for this field strength is $\partial^\mu {\cal
F}_{\mu\nu\ldots}=0$. We can now define the 3-form in the CY
originating from the 5-form ${\cal F}$ as an integral over the
2-cycle:
\begin{equation}
H_3= {1\over 2} \int_{S_2} {\cal F}_5= {\rm Im} \, \int_{S_2} {\cal
F}^+\ .
\end{equation}
Using the Hodge-decomposition for this 3-form (see Appendix C), we
can uniquely expand the real 3-form flux as follows:
\begin{eqnarray}\label{e3}
H_{3}={\mathcal{A}}\Omega
+\overline{\mathcal{B}}^{\bar{a}}\overline{\mathcal{D}}_{\bar{a}}\overline{\Omega}+
{\mathcal{B}}^{a}{\mathcal{D}}_{a}\Omega
+\overline{\mathcal{A}}\overline{\Omega}\ .\label{hodgeh3}
\end{eqnarray}
Here $\Omega=e^{K\over 2} \Omega^{3,0}(K_{6})$ is the covariantly
holomorphic three form of the Calabi-Yau and ${\mathcal{A}}$,
${\mathcal{B}}^{a}$,
 are the coefficients of the expansion. Our approach here is closely related to the one developed in
\cite{Denef:2004ze} and used for the derivation of the ``new
attractors'' in \cite{Kallosh:2005ax}. We have not included the
terms with the second derivative of $\Omega$ since  from the special
geometry of the moduli space \cite{deWit:1984pk} it is known that
${\mathcal{DD}}\Omega$ is not an independent form and, in fact, is
related to ${\mathcal{D}}\Omega$ through the Yukawa couplings
($G^{c\bar{d}}$ is the K\"{a}hler metric defined on $K_{6}$):
\begin{eqnarray}\label{e2}
{\mathcal{D}}_{a}{\mathcal{D}}_{b}\Omega=
iC_{abc}G^{c\bar{d}}\overline{{\mathcal{D}}}_{\bar{d}}\overline{\Omega}\
.
\end{eqnarray}

The next thing we need to do is to find the coefficients of the
expansion (\ref{hodgeh3}). Using the expression for the central
charge in terms of the covariantly holomorphic three form, namely
$Z=\int_{K_{6}} H_{3}\wedge\Omega$, we immediately find that
${\mathcal{A}}=i\overline{Z}$. Following the same procedure, we can
also obtain
${\mathcal{B}}^{a}=-iG^{a\bar{b}}\overline{{\mathcal{D}}}_{\bar{b}}\overline{Z}$.
Substituting these expressions into the expansion, we obtain (in
agreement with \cite{Ferrara:1996dd}, \cite{Kallosh:2005ax} and
\cite{Ferrara:1990dp}) the Hodge-decomposition of the 3-form flux
as:
\begin{eqnarray}\label{e4}
H_3= 2 {\rm Im} \left [  \,  Z \, \bar { \Omega}_3 +   \overline
{\mathcal{D}}_{ \bar a} \overline Z  G^{\bar a a}
 {\mathcal{D}}_{ a}  { \Omega}_3 \right ]= {\rm Im} \, \int_{S_2} {\cal F}^+ \
 .
\end{eqnarray}

Notice that this relation is valid  at any arbitrary point of the
moduli space. After integrating over the 3-cycles
 we can rewrite the above relation in
terms of the integer charges:
\begin{eqnarray}\label{e5}
h=2\mbox{Im}\Big[\ Z\overline{\Pi}+ G^{\bar{a}b}\overline
{\mathcal{D}}_{\bar a} \overline Z\ {{\mathcal{D}}}_{b}{\Pi}\ \Big]\
.
\end{eqnarray}
$h=(p^{\Lambda},q_{\Lambda})$ is the set of magnetic and electric
charges and $\Pi$ is the covariantly holomorphic period vector.
Clearly, at supersymmetric extrema (where ${\mathcal{D}}_{a}Z=0$),
this gives a simple (\cite{Ferrara:1995ih}-\cite{Moore:2004fg})
algebraic expression:
\begin{eqnarray}\label{e6}
h=2\mbox{Im}[Z\overline{\Pi}]\ .
\end{eqnarray}
\par
In more general case, we can use the minimization condition of the
effective potential of black hole $\partial V_{BH}=0$ which is
equivalent to (\cite{Ferrara:1997tw}):
\begin{eqnarray}\label{a21}
2({\mathcal{D}}_a Z) \bar Z + i C_{abc}G^{b\bar d} G^{c\bar e} \bar
{\mathcal{D}}_{\bar d} \bar Z \bar {\mathcal{D}}_{\bar e} \bar Z =0\
.
\end{eqnarray}
This readily gives:
\begin{eqnarray}\label{a22}
{\mathcal{D}}_a Z=-i\frac{C_{abc}G^{b\bar d} G^{c\bar e} \bar
{\mathcal{D}}_{\bar d} \bar Z \bar {\mathcal{D}}_{\bar e}\bar
Z}{2\bar{Z}}\ .
\end{eqnarray}
Using the special geometry relation ${\mathcal{D}}_{ a}
{\mathcal{D}}_b  Z= i C_{abc} G^{c\bar d} \bar {\mathcal{D}}_{\bar
d} \bar Z$ in the above equation and substituting into the result
for (\ref{e5}), we find the following equation:
\begin{eqnarray}\label{i4}
h=2\mbox{Im}\left [Z\overline{\Pi}- \frac{(\overline
{\mathcal{D}}_{\bar a}\overline {\mathcal{D}}_{\bar b} \overline Z)
G^{\bar a {c}}G^{\bar b {d} }{\mathcal{D}}_{{c}}{Z}
{{\mathcal{D}}}_{d} {\Pi}}{2{Z}}\right]\ .
\end{eqnarray}
The replacements made above are only valid in case that $Z\neq 0$.

This is the general form of the attractor equation, valid both for
supersymmetric and non-supersymmetric cases. We claim that solving
the above equation to find the moduli at horizon of the black hole
is equivalent to the minimization of the potential.

That (\ref{i4}) was derived using a type-IIB string theory
compactified on a CY manifold is somewhat auxiliary to the result.
It is an equation that is valid for the general case of 4d non-BPS
black holes (as defined in Section 2) in the framework of special
geometry. This is analogous to the situation that took place for the
BPS black hole attractor equation. While in \cite{Strominger:1996kf}
the equation $ h=2\mbox{Im}[Z\overline{\Pi}]$ was derived string
theory compactified on a CY manifold, in \cite{Ferrara:1996dd} the
derivation was extended to the general case of special geometry.
Thus our new non-BPS black hole attractor equation is valid for any
non-BPS black holes in special geometry, in particular when they are
derived from type-IIA string theory. In the next section, we
explicitly solve (\ref{i4}) for specific situations and find the
value of the moduli at the horizon of the black hole and compare our
results with minimizing the effective potential.
\par
Before we move on to the next section, we point out that care should
be taken with the precise meaning of (\ref{i4}). We use the symbol
${\mathcal{D}}$ to denote the fully covariant derivative:
\begin{eqnarray}\label{a23}
{\mathcal{D}}=\partial+x(\partial K)+\Gamma\ .
\end{eqnarray}
where $x$ denotes the K\"{a}hler weight of the object that the
covariant derivative acts on and $\Gamma$ denotes the Christoffel
symbol(s) of the Levi-Civita connection of the K\"{a}hler metric.
The K\"{a}hler weight $x$ of an object ${\mathcal{O}}$ is determined
by its transformation law under K\"{a}hler transformation. When
$K\rightarrow K+f(z) +\bar f(\bar z)$, the object  ${\mathcal{O}}$
transforms as ${\mathcal{O}} \rightarrow e^{ x( f(z) -\bar f(\bar
z))}{\mathcal{O}} $, see e. g. \cite{Kallosh:2000ve}. The K\"{a}hler
weights for the central charge $Z$ and its conjugate,  are
$\frac{1}{2}$ and $- \frac{1}{2}$ respectively. It follows from this
that the superpotential $W= e^{-K/2} Z$ has a covariant derivative
${\mathcal{D}} W = e^{- K/2}\left(\partial+{1\over 2}(\partial
K)\right) Z=
\partial W +\partial K W$ and the conjugate  has a vanishing
covariant derivative ${\mathcal{D}} \overline W = e^{-
K/2}\left(\partial-{1\over 2}(\partial K)\right) \overline Z=
\partial \overline W =0$. Notice that this definition of
covariant derivative is different from the derivative $\nabla$ used
in \cite{Tripathy:2005qp}, which does not include Christoffel
symbols. Of  course  the two operations are identical when acting on
scalar quantities (such as the effective potential and the
superpotential), but the distinction becomes important once we
consider second derivatives. Somewhat more detail on this issue
(especially with regards to the differences between the above quoted
expression for the extremal condition and that used in
\cite{Tripathy:2005qp}) can be found in Appendix B.

\section{Solving the Attractor Equation}

In this section, we consider three examples, explicitly solve the
attractor equation for them and extract the values of the moduli at
the horizon (the attractor point). The first of these are black hole
attractors in the framework of type-IIA string compactification,
where we take the internal space to be a Calabi-Yau manifold whose
volume is large. Next we consider a double-extremal example, in the
same background. In the final example, we study the mirror quintic
manifold in type-IIB, at the vicinity of the Gepner point. In each
case, we first solve the attractor equation directly to find the
moduli at horizon and then we compare our results with those from
the minimization of the effective potential presented in
\cite{Tripathy:2005qp}.

\subsection{Large Volume Calabi-Yau in the Absence of D6-brane}

Here we solve the attractor equation directly for black hole
attractors in the framework of type-IIA string theory
compactifications, in which the internal space is a Calabi-Yau
manifold with large volume. The BPS attractor equation for this
system was solved in \cite{Behrndt:1996jn} and the moduli at horizon
in the non-BPS case were found in \cite{Tripathy:2005qp} by
minimizing the effective potential of the black hole.

In the low-energy limit of ${\mathcal{N}}=2$ theory compactified on
a CY three-fold with $h^{1,1}(K_{6})=N$, the superpotential takes
the following form\footnote{Although it is not necessary to choose a
specific gauge, we work in the gauge $X^{0}=0$ for simplicity.}:
\begin{eqnarray}\label{a2}
W=q_{0}+q_{a}z^{a}-3D_{abc}p^{a}z^{b}z^{c}\ ,
\end{eqnarray}
in which $q_{0},q_{a},$ and $p^{a}$ are D0, D2 and D4-brane charges
respectively. For simplicity, we assume that there is no D6-brane in
this setup, $p^{0}=0$. The K\"{a}hler potential is then given by:
\begin{eqnarray}\label{a3}
K=-\ln\Big(-iD_{abc}\big(z^{a}-\bar{z}^{a}\big)
\big(z^{b}-\bar{z}^{b}\big)\big(z^{c} -\bar{z}^{c}\big)\Big)\ .
\end{eqnarray}
Also, we assume that there is no D2-brane\footnote{This condition
can be easily relaxed.}, namely $q_{a}=0$. Before jumping into the
task of calculating the terms which are involved in (\ref{i4}), it
is prudent to obtain the form of the solution. By solving
(\ref{i4}), we find $z^{a}$ as a function of charges and for the
present situation, the only charges are those associated with the
D0-brane and D4-brane, namely $q_{0}$ and $p^{a}$. Therefore, it is
quite clear that the most general symplectic vector which can be
constructed from $q_{0}$ and $p^{a}$ has the form
$z^{a}=ip^{a}t(q_{0},D)$, where we have defined
$D=D_{abc}p^{a}p^{b}p^{c}$. We can set all the $z^{a}$ to be purely
imaginary since the superpotential is quadratic with respect to
them. This implies that $t(q_{0},D)$ is a real function.

\noindent {\it First term of The Attractor Equation}

\noindent Now, we calculate the terms of r.h.s. of (\ref{i4}). We
know that:
\begin{eqnarray}\label{b1}
\ Z=e^{K/2}W \ ,\ \ \Pi=e^{K/2}\left(\begin{array}{c}
1 \\
z^a\\
F_0\\
F_{a} \\
\end{array}\right)\ .
\end{eqnarray}
Thus, considering the fact that the K\"{a}hler potential is real, we
can write the first term of the r.h.s. of (\ref{i4}) as:
\begin{eqnarray}\label{b2}
Z\overline{\Pi}=e^{K}W\left(\begin{array}{c}
1 \\
\bar{z}^a\\
\bar{F}_0\\
\bar{F}_{a} \\
\end{array}\right)\ .
\end{eqnarray}
$F_{a}=\partial_{a}F=\partial_{a}(D_{bcd}z^{b}z^{c}z^{d})=3D_{abc}z^{b}z^{c}$
and
$F_{0}=\partial_{0}F=\partial_{0}(\frac{D_{bcd}X^{b}X^{c}X^{d}}{X^0})=-D_{abc}z^{a}z^{b}z^{c}$.
Substituting the general form of $z^{a}=ip^{a}t$ into the above
expression, we get:
\begin{eqnarray}\label{b3}
&&\overline{F_a}=3D_{abc}(-ip^{b}t)(-ip^{c}t)=-3D_{a}t^{2}\ ,\\
&&\overline{F_{0}}=-D_{abc}(-ip^{a}t)(-ip^{b}t)(-ip^{c}t)=-iDt^{3}\
,
\end{eqnarray}
where $D_{a}=D_{abc}p^{b}p^{c}$. Using $M=D(2it)^{3}=-8iDt^{3}$ (see
(\ref{a11-3})), we easily find
$e^{K}=\frac{i}{M}=\frac{-1}{8Dt^{3}}$. Therefore, the first term of
the r.h.s. of (\ref{i4}) reads:
\begin{eqnarray}\label{b4}
2(Z\overline{\Pi})=-\frac{W}{4Dt^{3}}\left(\begin{array}{c}
1 \\
-ip^{c}t \\
-iDt^{3}\\
-3D_{c}t^{2}\\
\end{array}\right)\ .
\end{eqnarray}
Notice that $c$ is a free index and runs over 1 to $h^{1,1}=N$.
Substituting the general form of $z^{a}=ip^{a}t$ for (\ref{a2}), the
superpotential becomes $W=(q_{0}+3Dt^{2})$ --- this is clearly real.
Finally, the imaginary part of (\ref{b4}) gives the first term of
the r.h.s. of the attractor equation:
\begin{eqnarray}\label{a7}
2\mbox{Im}(Z\overline{\Pi})=\frac{W}{4Dt^{2}}\left(\begin{array}{c}
0 \\
p^{c} \\
Dt^{2}\\
0\\
\end{array}\right)\ .
\end{eqnarray}

\noindent  {\it Second Term of The Attractor Equation}

\noindent In order to evaluate the second term of the attractor
equation, we need to compute the first and second covariant
derivatives of the superpotential as well as the first covariant
derivative of the period vector.
\par
Following the notion of \cite{Tripathy:2005qp}, the covariant
derivative of the superpotential is:
\begin{eqnarray}\label{a12}
{\mathcal{D}}_{a}W=(\partial_{a}+(\partial_{a}K))W=-6D_{ab}z^{b}-\frac{3M_{a}}{M}W\
.
\end{eqnarray}
Here $D_{ab}=D_{abc}p^{c}$ and $M_{ab}$, $M_{a}$, and $M$ are
defined as:
\begin{eqnarray}
\label{a11-1}&&M_{ab}=D_{abc}(z^{c}-\bar{z}^{c})\ ,\\
\label{a11-2}&&M_{a}=D_{abc}(z^{b}-\bar{z}^{b})(z^{c}-\bar{z}^{c})\ ,\\
\label{a11-3}&&M=D_{abc}(z^{a}-\bar{z}^{a})(z^{b}-\bar{z}^{b})(z^{c}-\bar{z}^{c})\
.
\end{eqnarray}
If we again substitute the general form of $z^{a}=ip^{a}t$ into the
above expressions, we obtain $M_{ab}=2iD_{ab}t$,
$M_{a}=-4D_{a}t^{2}$, and $M=-8iDt^{3}$. Then, the covariant
derivative of the superpotential reads:
\begin{eqnarray}\label{a13}
{\mathcal{D}}_{a}W=-3itD_{a}\Big(2-\frac{W}{2Dt^{2}}\Big)\ \ \ ,\ \
\
\overline{\mathcal{D}}_{\bar{a}}\overline{W}=+3itD_{a}\Big(2-\frac{W}{2Dt^{2}}\Big)\
.
\end{eqnarray}
The second covariant derivative of the superpotential is:
\begin{eqnarray}\label{a8}
{\mathcal{D}}_{a}{\mathcal{D}}_{b}W=(\partial_a+(\partial_{a}K)){\mathcal{D}}_bW-\Gamma^d_{ab}{\mathcal{D}}_dW\
,
\end{eqnarray}
with the Christoffel symbols of the K\"{ahler} metric given by
$\Gamma^{d}_{ab}=G^{d\bar{e}}\partial_{a}G_{\bar{e}b}$. Expressions
for the metric and inverse metric can be found in
\cite{Tripathy:2005qp}:
\begin{eqnarray}\label{a9}
G_{a\bar{b}}&=&\frac{3}{M}\left(2M_{ab}-\frac{3}{M}M_aM_b\right)\\
G^{a\bar{b}}&=&\frac{M}{6}\left(M^{ab}-\frac{3}{M}(z^{a}-\bar{z}^{a})(z^{b}-\bar{z}^{b})\right)\
.
\end{eqnarray}
These, in turn, imply that the Christoffel symbols take the
following form:
\begin{eqnarray}\label{a10}
\Gamma^{d}_{ab}=M^{de}D_{eab}-\frac{3}{M}\Big(M_a\delta^d_b+M_b\delta^d_a-M_{ab}(z^d-\bar{z}^d)\Big)\
.
\end{eqnarray}
Combining (\ref{a13}) and (\ref{a10}), we get:
\begin{eqnarray}\label{a14}
\Gamma^{d}_{ab}{\mathcal{D}}_{d}W=-12D_{ab}+18\frac{D_{a}D_{b}}{D}+\frac{3W}{2Dt^{2}}
\big(2D_{ab}-3\frac{D_{a}D_{b}}{D}\big)\ .
\end{eqnarray}
We can also compute the first piece of (\ref{a8}):
\begin{eqnarray}\label{a15}
(\partial_a+(\partial_{a}K)){\mathcal{D}}_bW=-6D_{ab}+18\frac{D_{a}D_{b}}{D}+\frac{3W}{2Dt^{2}}
\big(D_{ab}-3\frac{D_{a}D_{b}}{D}\big)\ .
\end{eqnarray}
After all this, we obtain the second covariant derivative of the
superpotential:
\begin{eqnarray}\label{a16}
{\mathcal{D}}_{a}{\mathcal{D}}_{b}W=3D_{ab}\Big(2-\frac{W}{2Dt^{2}}\Big)\
.
\end{eqnarray}
\par
The last thing we need to compute is the covariant derivative of the
period vector. Using the expression for the period vector
(\ref{b1}), we now find its covariant derivative:
\begin{eqnarray}\label{a17}
{\mathcal{D}}_{d}\Pi=e^{K/2}{\mathcal{D}}_{d}\left(
\begin{array}{c}
1 \\
z^{c} \\
F_0 \\
F_{c} \\
\end{array} \right)=e^{K/2}\left(\begin{array}{c}
\partial_{d}K \\
\delta^{c}_{d}+\partial_{d}K\ z^{c} \\
-3D_{def}z^ez^f-\partial_{d}KD_{efh}z^ez^fz^h \\
6D_{cde}z^{e}+3\partial_{d}K\ D_{cef}z^{e}z^{f} \\
\end{array}\right)\ .
\end{eqnarray}
We can easily see that
$\partial_{d}K=-3\frac{M_{d}}{M}=\frac{3iD_{d}}{2Dt}$. Substituting
this result in the above expression, we get:
\begin{eqnarray}\label{c9}
{\mathcal{D}}_{d}\Pi=e^{K/2}\left(\begin{array}{c}
\frac{3}{2}i\frac{D_d}{D}\frac{1}{t} \\
\delta^{c}_{d}-\frac{3}{2}p^{c}\frac{D_{d}}{D} \\
\frac{3}{2}D_{d}t^{2} \\
6iD_{cd}t-\frac{9}{2}i\frac{D_{d}D_{c}}{D}t \\
\end{array} \right)\ .
\end{eqnarray}
Combining the above result with (\ref{a13}), (\ref{a9}) and
(\ref{a16}), we get the second term of the attractor equation:
\begin{eqnarray}\label{a18}
2\mbox{Im}\Bigg[\frac{(\overline{{\mathcal{D}}}_{\bar{a}}\overline{{\mathcal{D}}}_{\bar{b}}\bar{Z})
G^{\bar{a}c}G^{\bar{b}d}{\mathcal{D}}_{c}Z
{\mathcal{D}}_{d}\Pi}{2Z}\Bigg]=-\frac{Dt^{2}}{2W}(-2+\frac{W}{2Dt^{2}})(4-\frac{W}{Dt^{2}})
\left(
  \begin{array}{c}
    0 \\
    -p^{c}\\
    3Dt^{2}\\
    0 \\
  \end{array}
\right)\ .
\end{eqnarray}

\noindent {\it  Adding the Two Terms Up}

\noindent Now that we have the two terms of the r.h.s. of the
attractor equation, we can form the equation and solve it in order
to find $t(q_{0},D)$. We know $h=(p^{\Lambda},q_{\Lambda})$ is the
set of charges and for this setup we have:
\begin{eqnarray}\label{d4} h=\left(\begin{array}{c}
0 \\
p^{c} \\
q_0 \\
0 \\
\end{array}
\right)\ .
\end{eqnarray}
If we define the new variable $Y=\frac{W}{Dt^{2}}$, then we can
write the attractor equation (\ref{i4}) in the following way:
\begin{eqnarray}\label{a19}
\left(
  \begin{array}{c}
    0 \\
    p^{c} \\
    (Y-3)Dt^{2}\\
    0 \\
  \end{array}
\right)= \left(
\begin{array}{c}
0 \\
\Big(\frac{1}{4}Y-\frac{1}{2Y}\big(-2+\frac{1}{2}Y\big)(4-Y)\Big)p^{c} \\
\Big(\frac{1}{4}Y+\frac{3}{2Y}\big(-2+\frac{1}{2}Y\big)(4-Y)\Big)Dt^{2} \\
0 \\
\end{array}
\right)\ ,
\end{eqnarray}
in which we have substituted $q_{0}$ by
$q_{0}=W-3Dt^{2}=(Y-3)Dt^{2}$. Now, we find that both non-vanishing
rows of the above matrix equation lead to the following equation for
$Y$:
\begin{eqnarray}\label{a20}
(Y-4)(Y-2)=0\ .
\end{eqnarray}
The solution $Y=4$, which corresponds to the supersymmetric solution
(because ${\mathcal{D}}_{a}W=0$ at $Y=4$, see (\ref{a13})), leads us
to the result $t=\sqrt{\frac{q_{0}}{D}}$. However, the solution
$Y=2$, corresponding to the non-supersymmetric solution
(${\mathcal{D}}_{a}W\neq 0$ at $Y=2$), gives us
$t=\sqrt{-\frac{q_{0}}{D}}$.

\noindent {\it Minimization of The Effective Potential}

\noindent In the previous section, we explicitly solved the
attractor equation and found the moduli at the horizon (the
attractor point) of the black hole. In \cite{Tripathy:2005qp},
however, the moduli at horizon is found by the minimization of the
effective potential of the black hole. Here we see that these
approaches are equivalent.

Minimization of the effective potential leads to the following
equation\footnote{The form of this equation in
\cite{Tripathy:2005qp} is slightly different from that presented
here. But in Appendix B, we show that they are equivalent.}:
\begin{eqnarray}\label{a4}
\partial_{a}V_{eff}={\mathcal{D}}_{a}\Big(|Z|^{2}+|{\mathcal{D}}Z|^{2}\Big)=
e^{K}\Big(g^{b\bar{c}}({\mathcal{D}}_{a}{\mathcal{D}}_{b}W)
\bar{{\mathcal{D}}}_{\bar{c}}\overline{W}+2({\mathcal{D}}_{a}W)\overline{W}\Big)=0\
.
\end{eqnarray}
Using the general form of $z^{a}=ip^{a}t$, the equation governing
$t$ takes the following form:
\begin{eqnarray}\label{a6}
(q_{0}-t^{2}D)(q_{0}+t^{2}D)=0\ .
\end{eqnarray}
The solution $t_{1}=\sqrt{\frac{q_{0}}{D}}$ is the supersymmetric
one (where ${\mathcal{D}}_{a}W$ vanishes) and
$t_{2}=\sqrt{-\frac{q_{0}}{D}}$ is the non-supersymmetric one (where
${\mathcal{D}}_{a}W\neq 0$). Thus we confirm that the two procedures
lead to the same answer.

\subsection{An Example of the Non-BPS Double-extremal Black Hole}

In this subsection, we will first review  an example of a
double-extremal black hole given in \cite{Behrndt:1996jn}. It
corresponds to a special case of the model studied in Sec. 5.1 when
there are only 3 moduli coordinates $z^1, z^2, z^3$. The only
non-vanishing component of $D_{abc}$ is $D_{123}= -1/6$ and
therefore $D= -p^1 p^2 p^3$.
\par
As explained in Section 3, double-extremal black holes are
Reissner-Nordstr\"{o}m type solutions of the theory defined by
action (\ref{scalaraction2}) where the moduli fields $z^a, \bar
z^{\bar{a}}$ take constant values. The values of the moduli at the
horizon of the black hole are determined by solving the attractor
equation (or equivalently by finding the critical points of the
black hole potential). Therefore, the constant values for
double-extremal black holes are defined by the solution of the
attractor equation.

\noindent {\it Double-extremal BPS Black Hole}

\noindent As shown in \cite{Kallosh:1996tf}-\cite{Behrndt:1996jn}
the 4-dimensional metric, in the double-extremal limit, defines an
extreme Reissner-Nordstr\"{o}m metric $ ds_{4}^{2}=e^{2U} \ dt^{2}
-e^{-2U} d\vec{x}^{2} $, where:
\begin{equation}
 e^{-2U}= \sqrt{H_{0}D_{abc}H^{a}H^{b}H^{c}}= \Big(1+{M\over r}\Big)^2 \ , \qquad M^2= 2 \sqrt{q_0
 D}\ .
\label{metric}
\end{equation}
Here the harmonic (i.e. $\partial_i
\partial_i H_0= \partial_i \partial_i H^a=0$) functions are given
by:
\begin{eqnarray}\label{Doub-Ext3}
&&H_{0}=\sqrt{2}q_{0}\Big(k+\frac{1}{r}\Big)\ , \ \ \
H^{1}=-\sqrt{2}p^{1}\Big(k+\frac{1}{r}\Big)\ ,\nonumber\\
&&H^{2}=\sqrt{2}p^{2}\Big(k+\frac{1}{r}\Big)\ , \ \ \
H^{3}=\sqrt{2}p^{3}\Big(k+\frac{1}{r}\Big)\ .
\end{eqnarray}
$k^{-4}=-4q_{0}p^{1}p^{2}p^{3}$ and $r=|\vec{x}|$. It is assumed
that $q_0 D = -q_{0}p^{1}p^{2}p^{3}>0$. The constant values of the
three moduli are:
\begin{eqnarray}\label{Doub-Ext4}
&&z^{1}=i\sqrt{\frac{H_{0}H^{1}}{H^{2}H^{3}}}=i \, p^1
\sqrt{\frac{q_{0}}{D}}\
,\\
&&z^{2}=i\sqrt{\frac{H_{0}H^{2}}{H^{1}H^{3}}}=i \, p^2
\sqrt{\frac{q_{0}}{D}}\
,\\
&&z^{3}=i\sqrt{\frac{H_{0}H^{3}}{H^{1}H^{2}}}=i \,  p^3
\sqrt{\frac{q_{0}}{D}}\ .
\end{eqnarray}
It is evident that all of the above only depend on the electric
charge $q_{0}$ and the magnetic charges $p^{1}$, $p^{2}$, and
$p^{3}$. Moreover, as mentioned in Section 3, the solution for the
four vector fields is given by (\ref{vector}). For this example, we
need to introduce the values of the dressed charges and of $M$. To
do this, one first has to find the vector couplings. Using
(\ref{vectorcoupling}) we obtain ($t= \sqrt {q_0/D}$):
\begin{eqnarray}\label{N-coupling}
{\mathcal{N}}_{\Lambda\Sigma}=\left(
                                \begin{array}{cc}
                                  iDt^{3} & 0 \\
                                  0 & -6iD_{ab}t+9it\frac{D_{a}D_{b}}{D} \\
                                \end{array}
                              \right)\ .
\end{eqnarray}
The fact that the above matrix has no real entries implies that we
do not have any axions for this example, as expected. Using
(\ref{dressed}) we simply find the values of the dressed electric
and magnetic charges as $P^0=0$, $P^a= {1\over 2}p^a$, $Q^0= {1\over
2 t}$ and $Q^a=0$. Also we have $M^2= 2 \sqrt{q_0 D}= -2 (\mbox{Im}
\, {\cal N}_{\Lambda \Sigma })(Q^\Lambda Q^\Sigma + P^\Lambda
P^\Sigma)$. In addition to the constant values of the scalar fields,
one also has to find the values of the the vector fields. Using
(\ref{vector}), the electric and magnetic fields are given in terms
of harmonic functions $H_{0}$, $H^{1}$, $H^{2}$, and $H^{3}$ as:
\begin{equation}
{\cal F}^0_{i0}=  \partial_i\psi^0\ , \qquad {\cal G}_{a\, i0}=
\partial_i\chi_a\ . \label{elmag}
\end{equation}
The electric and magnetic potentials are proportional to the
inverses of the harmonic functions:
\begin{eqnarray}\label{pot}
\psi^0=\frac{1}{\sqrt{2}}(H_0)^{-1}\ ,\qquad
\chi_a=\frac{1}{\sqrt{2}}(H^a)^{-1} \ .
\end{eqnarray}

\noindent {\it Double-extremal Non-BPS Black Hole}

\noindent Now we wish to consider the double-extremal black hole for
the non-supersymmetric case. In the previous section, we explicitly
solved the attractor equation (\ref{i4}) to find the values of
moduli coordinates at the horizon of the black hole. In the general
setup, we had $N$ moduli coordinates, an electric charge $q_{0}$ and
magnetic charges $p^{a}$. Comparing supersymmetric and non
supersymmetric solutions\footnote{The supersymmetric solution is
$z^{a}=ip^{a}\sqrt{\frac{q_{0}}{D}}$ whereas the non-supersymmetric
solution is given by $z^{a}=ip^{a}\sqrt{-\frac{q_{0}}{D}}$.}, we
immediately realize that in order to obtain the double-extremal
black hole in the non-BPS case one only needs to substitute $q_{0}$
by $-q_{0}$ in the supersymmetric double-extremal black hole
solution with $-q_0 D>0$.

This leads to the following expressions for the harmonic functions:
\begin{eqnarray}\label{nonbps}
&&\tilde H_{0}=-\sqrt{2}q_{0}\Big(\tilde{k}+\frac{1}{r}\Big)\ , \ \
\
\tilde H^{1}=-\sqrt{2}p^{1}\Big(\tilde{k}+\frac{1}{r}\Big)\ ,\nonumber\\
&&\tilde H^{2}=\sqrt{2}p^{2}\Big(\tilde{k}+\frac{1}{r}\Big)\ , \ \ \
\tilde H^{3}=\sqrt{2}p^{3}\Big(\tilde{k}+\frac{1}{r}\Big)\ ,
\end{eqnarray}
where, in this case, $\tilde{k}^{-4}=+4q_{0}p^{1}p^{2}p^{3}$. In
terms of these harmonic functions the metric is the same as in
(\ref{metric}). The scalars, again in terms of new harmonic
functions, are the same as in
 (\ref{Doub-Ext4}). This corresponds to changing $q_{0}$ to
$-q_{0}$ and we get $\tilde{M}^2= 2\sqrt {-q_0 D}$.

Therefore, considering the BPS solution (\ref{Doub-Ext4}), the
constant values of the three moduli coordinates for the non-BPS
black hole are given by:
\begin{eqnarray}\label{Doub-ExtN}
&&\tilde{z}^{1}=i\sqrt{\frac{\tilde H_{0}\tilde H^{1}}{\tilde
H^{2}\tilde H^{3}}}=i \, p^1 \sqrt{\frac{-q_{0}}{D}}\
,\\
&&\tilde{z}^{2}=i\sqrt{\frac{\tilde H_{0}\tilde H^{2}}{\tilde
H^{1}\tilde H^{3}}}=i \, p^2 \sqrt{\frac{-q_{0}}{D}}\
,\\
&&\tilde{z}^{3}=i\sqrt{\frac{\tilde H_{0}\tilde H^{3}}{\tilde
H^{1}\tilde H^{2}}}=i \, p^3 \sqrt{\frac{-q_{0}}{D}}\ .
\end{eqnarray}
From (\ref{pot}) we can also find the values of the electric and
magnetic potentials; in terms of harmonic functions these are:
\begin{equation}
\tilde {\cal F}^0_{i0}=  \partial_i\tilde \psi^0, \qquad \tilde
{\cal G}_{a\, i0}=  \partial_i\tilde{\chi}_a \ .\label{elmag}
\end{equation}
The electric and magnetic potentials are related to the inverse of
harmonic functions in the following way:
\begin{equation}
\tilde \psi^0=\frac{1}{\sqrt{2}}(\tilde H_0)^{-1} \qquad
\tilde{\chi}_a=\frac{1}{\sqrt{2}}(\tilde H^a)^{-1} \ .\label{pot2}
\end{equation}

\subsection{Mirror Quintic}
In this section, we consider black holes in the framework of
type-IIB string theory compactified on mirror quintic Calabi-Yau
3-folds. We closely follow the notation of \cite{Tripathy:2005qp}
and \cite{Giryavets:2003vd}. For a quintic hypersurface in $P^{4}$,
the mirror quintic $M$ is obtained by the following
quotient(\cite{Candelas:1990rm}):
\begin{eqnarray}\label{ff0}
M=\Big(\sum_{i=1}^{5}Z_{i}^{5}-5\psi\prod_{i=1}^{5}Z_{i}\Big)\Big/(Z_{5})^{3}\
,
\end{eqnarray}
where $\psi$ is a complex coefficient. The Hodge numbers of the
mirror quintic Calabi-Yau are $h^{1,1}(M)=101$ and $h^{2,1}(M)=1$.
In IIB theory, the vector multiplet moduli space which corresponds
to the deformations of the complex structure is parameterized by
$\psi$ and is one dimensional. In terms of homology there are, in
general, $2(h^{2,1}(M)+1)$ nontrivial 3-cycles $\{A^{a},B_{a}\}$.
From the holomorphic 3-form $\Omega$ of the Calabi-Yau manifold, we
can find the special coordinates $z^{a}$ of the vector multiplet
moduli space and the prepotential $F$:
\begin{eqnarray}\label{ff5}
\int_{A^{a}}\Omega=z^{a}\ \ ,\ \ \int_{B_{a}}\Omega=\partial_{a}F\ .
\end{eqnarray}

The K\"{a}hler potential and the superpotential are then given by:
\begin{eqnarray}\label{ff6}
K=-\log(-i\Pi^{\dag}\cdot\Sigma\cdot\Pi)\ \ ,\ \
W=h^{T}\cdot\Sigma\cdot\Pi\ ,
\end{eqnarray}
where $\Sigma$ is a matrix given in Appendix D, $h=(p^{a},q_{a})$ is
the set of electric and magnetic charges and $\Pi$ is the period
function. We notice that both $h$ and $\Pi$ are four dimensional
column vectors. Now, we want to solve the attractor equation
((\ref{i4})) for the case of mirror quintic in the vicinity of the
Gepner point ($\psi=0$) at which the holomorphic 3-form of the
Calabi-Yau is well known as a power series (\cite{Candelas:1990rm}).
In this limit, we find all ingredients of the attractor equation as
a series in terms of $\psi$ and then we only keep linear terms. The
K\"{a}hler potential and the superpotential are then expressed as:
\begin{eqnarray}\label{f0}
&&K=C_{0}-\log\Big(1+(2-\sqrt{5})\frac{c_{1}^{2}}{c_{0}^{2}}|\psi|^{2}-
(2-\sqrt{5})\frac{c_{2}^{2}}{c_{1}^{2}}|\psi|^{4}+\cdots\Big)\ ,\\
&&W=\frac{1}{25}\Big(\frac{2\pi i}{5}\Big)^{3}\Big(c_{0}n\cdot
p_{0}+c_{1}n\cdot p_{1}\psi+c_{2}n\cdot p_{2}\psi^{2}+\cdots\Big)\ .
\end{eqnarray}
The constants $C_{0}$, $c_{i}$ and the column vectors $p_{i}$ are
defined in Appendix D. $n$ is a column vector which is related to
the charges by $n=5\tilde{m}h$, where matrix $\tilde{m}$ can be
found in appendix [C]. From the K\"{a}hler potential, it is
straightforward to calculate the K\"{a}hler metric and the
Christoffel symbols of the Levi-Civita connection associated with
this metric:
\begin{eqnarray}\label{f2}
&&G^{\psi\bar{\psi}}=-\frac{c_{0}^{2}}{c_{1}^{2}(2-\sqrt{5})}\Bigg[1+\Big(2(2-\sqrt{5})
\frac{c_{1}^{2}}{c_{0}^{2}}+4\frac{c_{2}^{2}}{c_{1}^{2}}\Big)|\psi|^{2}+\cdots\Bigg]\
,\\
&&\Gamma^{\psi}_{\psi\psi}=-2\Big(\
(2-\sqrt{5})\frac{c_{1}^{2}}{c_{0}^{2}}+2\frac{c_{2}^{2}}{c_{1}^{2}}\
\Big)\bar{\psi}\ .
\end{eqnarray}
Now we are able to compute the ingredients of the attractor
equation. The covariant derivatives of the superpotential are given
by:
\begin{eqnarray}\label{f1}
&&{\mathcal{D}}_{\psi}W=\frac{1}{25}\Big(\frac{2\pi
i}{5}\Big)^{3}\Big(c_{1}n\cdot p_{1}+2c_{2}n\cdot
p_{2}\psi-\frac{c_{1}^{2}}{c_{0}^{2}}(2-\sqrt{5})c_{0}n\cdot
p_{0}\bar{\psi}\Big)\ ,\\
&&{\mathcal{D}}_{\psi}{\mathcal{D}}_{\psi}W=\frac{2}{25}\Big(\frac{2\pi
i}{5}\Big)^{3}\Big(c_{2}n\cdot p_{2}+3c_{3}n\cdot
p_{3}\psi+2\frac{c_{2}^{2}}{c_{1}}n\cdot p_{1}\bar{\psi}\Big)\ .
\end{eqnarray}
The period vector in terms of $\psi$ is:
\begin{eqnarray}\label{f4}
\Pi=\frac{1}{5}\Big(\frac{2\pi
i}{5}\Big)^{3}\Big(c_{0}\tilde{m}p_{0}+c_{1}\tilde{m}p_{1}\psi+c_{2}\tilde{m}p_{2}\psi^{2}+\cdots
\Big)\ .
\end{eqnarray}
For the covariant derivative of the period vector, we get:
\begin{eqnarray}\label{f5}
{\mathcal{D}}_{\psi}\Pi=\frac{1}{5}\Big(\frac{2\pi
i}{5}\Big)^{3}\Big(c_{1}\tilde{m}p_{1}+2c_{2}\tilde{m}p_{2}\psi-(2-\sqrt{5})
\frac{c_{1}^{2}}{c_{0}}\tilde{m}p_{0}\bar{\psi}\Big)\ .
\end{eqnarray}
So far, we have computed all the ingredients of the r.h.s. of the
attractor equation (\ref{i4}) for the mirror quintic. The only thing
we need is the l.h.s. of (\ref{i4}), namely the charges which should
be expressed in terms $n$. This is given by:
\begin{eqnarray}\label{f7}
h=\left(
\begin{array}{c}
p^{a} \\
q_{a} \\
\end{array}\right)
=\frac{1}{5}\left(
\begin{array}{c}
-20n_{1}\\
2(4n_{1}-n_{2})\\
11n_{1}\\
-4n_{1}+n_{2}\\
\end{array}
\right)\ ,
\end{eqnarray}
in which we assumed that vector $n$ has the form
$n=(n_{1},n_{2},n_{2},n_{1})$. This assumption is not necessary but
it makes the calculations easier. Specifically, one can see that
such an assumption ensures all the $n\cdot p_{i}$ are real. Next we
form the attractor equation and ignore all quadratic and higher
order terms in $\psi$. Then, we have:
\begin{eqnarray}\label{f6}
2\mbox{Im}\Bigg[ Z\overline{\Pi}-(G^{\psi\bar{\psi}})^{2}
\frac{(\overline{{\mathcal{D}}}_{\bar{\psi}}\overline{{\mathcal{D}}}_{\bar{\psi}}{\overline{Z}})
{\mathcal{D}}_{\psi}Z{\mathcal{D}}_{\psi}\Pi}
{2Z}\Bigg]-h=0=N_{1}+N_{2}\psi\ .
\end{eqnarray}
$N_{1}$ is expressed in the following way:
\begin{eqnarray}\label{f81}
&&N_{1}=-h+\frac{2}{5\sqrt{2+2\sqrt{5}}}\ \tilde{m}\Big((n\cdot
p_{0})\mbox{Im}\bar{p}_{0}-\frac{1}{(2-\sqrt{5})^{2}}\frac{c_{0}c_{2}}{c_{1}^{2}}\frac{(n\cdot
p_{1})^{2}}{n\cdot p_{0}}\mbox{Im}p_{1}\Big)\ ,
\end{eqnarray}
For $N_{2}$ we obtain (assuming $\psi$ is real):
\begin{eqnarray}\label{f82}
N_{2}=\frac{2}{5\sqrt{2+2\sqrt{5}}}\
\tilde{m}&\Bigg[&\frac{c_{1}}{c_{0}}\Big((n\cdot
p_{1})\mbox{Im}\bar{p}_{0}+(n\cdot
p_{0})\mbox{Im}\bar{p}_{1}\Big)-\frac{2}{(2-\sqrt{5})^{2}}\frac{c_{0}c_{2}^{2}}{c_{1}^{3}}\frac{(n\cdot
p_{1})^{2}}{n\cdot
p_{0}}\mbox{Im}(p_{1}+p_{2})\nonumber\\
&&+\frac{1}{(2-\sqrt{5})^{2}}\frac{c_{2}}{c_{1}}\frac{(n\cdot
p_{1})^{3}}{(n\cdot
p_{0})^{2}}\mbox{Im}p_{1}-\frac{3}{(2-\sqrt{5})^{2}}\frac{c_{0}c_{3}}{c_{1}^{2}}(n\cdot
p_{1})\mbox{Im}p_{1}\nonumber\\
&&+\frac{1}{(2-\sqrt{5})}\frac{c_{2}}{c_{1}}\Big((n\cdot
p_{1})\mbox{Im}p_{1}+\frac{(n\cdot p_{1})^{2}}{n\cdot
p_{0}}\mbox{Im}p_{0}\Big)\Bigg]\ .
\end{eqnarray}
In order to find a solution, we need to pick a specific set of
charges up such that $N_{1}/N_{2}\ll1$. Therefore, we set $N_{1}=0$
and this equation determines the ratio $\frac{n_{1}}{n_{2}}$. The
equation we find for the ratio $\frac{n_{1}}{n_{2}}$ is:
\begin{eqnarray}\label{ff7}
\Big(\frac{n_{1}}{n_{2}}\Big)^{2}-0.063\Big(\frac{n_{1}}{n_{2}}\Big)-0.121=0\
.
\end{eqnarray}
The above equation clearly has two solutions:
$\Big(\frac{n_{1}}{n_{2}}\Big)=0.381$, which corresponds to the
supersymmetric solution (it is easy to see that
${\mathcal{D}}_{\psi}W$ vanishes for this solution because $n\cdot
p_{1}=0$ for $\Big(\frac{n_{1}}{n_{2}}\Big)=0.381$) and
$\Big(\frac{n_{1}}{n_{2}}\Big)=-0.319$, which corresponds to the
non-supersymmetric solution (${\mathcal{D}}_{\psi}W\neq 0$ for
$\Big(\frac{n_{1}}{n_{2}}\Big)=-0.319$). This result is in agreement
with the one which is obtained by minimization of the black hole
potential in \cite{Tripathy:2005qp}.

\section{Discussion}

In analyzing the non-BPS black hole attractor mechanism for we have
provided a reasonable physical argument as to why such a mechanism
depends on the extremality (in the sense of zero temperature) of the
black hole. Whilst not entirely rigorous, such thinking may be
valuable as we seek to understand other situations in which we may
have attractors; for example, flux vacua.

We have also developed an explicit form of the black hole attractor
equation; deriving it from a minimization condition on $V_{BH}$ and
the Hodge decomposition of the 3-form flux. The veracity of this
equation has been demonstrated for a number of the examples. In
these examples when a non-supersymmetric minimum exists, it does so
in isolation --- this is a point which we develop further below.

There is an apparent conceptual similarity between non-BPS extremal
black holes and the O'Raifeartaigh model of spontaneous
supersymmetry breaking. There are certain superpotentials which do
not admit a supersymmetric minimum of the potential but do admit a
non-supersymmetric one. A classic example is the O'Raifeartaigh
model (\cite{O'Raifeartaigh:1975pr}) for 3 scalars: $W=\lambda
\Phi_1(\Phi^2_3- M^2) +\mu \Phi_2 \Phi_3$. Here $\partial_1 W=
\lambda (\Phi^2_3- M^2)$, $\partial_2 W= \mu \Phi_3$ and $\partial_3
W= 2\lambda \Phi_1 \Phi_3+\mu \Phi_2$. All 3 derivatives cannot
simultaneously vanish. However, if $M_2< \mu^2/2\lambda^2$, the
potential has an absolute minimum with positive value at $\Phi_2
=\Phi_3=0$, along with a flat direction in the $\Phi_1$ direction.

In models of the above type the system cannot decay to a
supersymmetric ground state since such a state does not exist, so
the non-SUSY vacuum is stable. The same is true of the non-BPS black
hole --- there is a choice of fluxes which leads to an effective
superpotential such that $V_{BH}$ does not  admit a supersymmetric
minimum of the potential but does admit a non-supersymmetric one.
Actually, there are also some additional constraints we should
apply. It is not enough to show that no spherically-symmetric BPS
state exists, we should also confirm that there are no
multi-centered BPS solutions. Such solutions are discussed (with
explicit examples) in \cite{Bates:2003vx}. Although in the cases we
have studied such multi-centered solutions do not exist, it is clear
that this will not always be true, and in addition to checking that
we have an attractive non-BPS solution, we must ensure that the
chosen charges do not also allow multiple states.

In our examples with an appropriate choice of fluxes the non-BPS
black holes are stable; there is no way to get a supersymmetric
black hole or BPS black hole composites for the given set of fluxes
and, furthermore, since $T=0$ they do not evaporate. It would be
interesting to develop a more general strategy on finding such
stable non-BPS black holes as, so far, there are just a few known
examples.

Intriguingly there is a similarity between the non-BPS attractor
equations for black holes and those for flux vacua that were studied
in \cite{Kallosh:2005ax}. This similarity suggests that the
attractor mechanism could be successful in providing a realization
of an effective way of achieving stable SUSY-breaking. Some recent
analysis of this issue in \cite{Intriligator:2005aw} shows just how
difficult it may be to avoid a runaway of the non-SUSY vacua into
the SUSY ones. In the case of O'Raifeartaigh model and the non-BPS
black holes, however, this runaway does not take place since there
is no supersymmetric vacua. It thus remains a challenge to construct
the analog of the stable non-BPS extremal black holes in dS flux
vacua.


\ \leftline{\bf Acknowledgments}

We are  particularly grateful to S. Ferrara, F. Denef,  A.
Giryavets,  J. Hsu,   and A. Linde for valuable discussions of flux
vacua and attractors. This work is supported by NSF grant 0244728.
N.S. and M.S. are also supported by the U.S. Department of Energy
under contract number DE-AC02-76SF00515.

\appendix
\section{Deriving the Effective Potential}
The derivation of (\ref{lagr}) and (\ref{constr1}) from
(\ref{scalaraction2}) using the metric ansatz (\ref{ansatz}) is
straightforward, but calculationally involved. Here we highlight the
salient points. Our starting point is the Einstein-Maxwell action
quoted in Section 2:
\begin{equation}
-{R\over 2} + G_{a\bar a} \partial_ \mu z^a   \partial_\nu \bar
z^{\bar a} g^{\mu\nu} +
 \frac{1}{2}\mu_{\Lambda \Sigma} {\cal F}^{\Lambda}_{\mu \nu}  {\cal
F}^{ \Sigma}_{\lambda \rho}  g^{\mu \lambda} g^{\nu \rho} +
\frac{1}{2}\nu_{\Lambda \Sigma} {\cal F}^{\Lambda}_{\mu
\nu}\left(\ast{\cal F}^{ \Sigma}_{\lambda \rho}\right) g^{\mu
\lambda} g^{\nu \rho}
 \ .
\label{scalaraction3}
\end{equation}
We have used $\mu$ and $\nu$ instead of the imaginary and real parts
of $\cal{N}$ for clarity. We define the dual through:
\begin{equation}
(*F)_{\mu\nu}=\frac{1}{2}\sqrt{-g}\epsilon_{\mu\nu\rho\sigma}F^{\rho\sigma}\
,
\end{equation}
with $\epsilon_{t\tau\theta\phi}=1=-\epsilon^{t\tau\theta\phi}$.
Using the metric ansatz (\ref{ansatz}) and multiplying by
$\sqrt{-g}$ we can obtain the Lagrangian terms. The gravity term
gives:
\begin{eqnarray}
-\sqrt{-g}\
\frac{R}{2}&=&-\frac{1}{2}\left(\frac{c^4e^{-2U}\sin(\theta)}{\sinh^4(c\tau)}\right)
\left(-\frac{2e^{2U}\sinh^4(c\tau)\left(c^2-U'(\tau)^2+U''(\tau)\right)}{c^4}\right)\nonumber\\
\ &=&-\sin(\theta)\left(\frac{\partial U}{\partial\tau}\right)^2\ .
\end{eqnarray}
We have dropped the constant and the $U''(\tau)$ terms (the latter
is a total derivative). Similarly, the scalar kinetic term can be
readily calculated to give:
\begin{equation}
\sqrt{-g}\ G_{a\bar a} \partial_ \mu z^a   \partial_\nu \bar z^{\bar
a} g^{\mu\nu}=-\sin(\theta)G_{a\bar a} \partial_{\tau} z^a
\partial_{\tau} \bar z^{\bar a}\ .
\end{equation}
The vector terms are somewhat trickier. First we define:
\begin{equation}
\mathcal{G}^{\mu\nu}_{\Sigma}=-i\mu_{\Sigma\Lambda}\left(*\mathcal{F}\right)^{\Lambda\,\mu\nu}
-\nu_{\,\Sigma\Lambda}\mathcal{F}^{\Lambda\,\mu\nu}\ .\label{G}
\end{equation}
Then we define electric and magnetic potentials through:
\begin{eqnarray}
\mathcal{F}^{\Sigma}_{t\tau}&=&\partial_{\tau}\psi^{\Sigma}\nonumber\\
\mathcal{G}_{\Sigma\,t\tau}&=&\partial_{\tau}\chi_{\Sigma}\
.\label{empot}
\end{eqnarray}
We can then use (\ref{G}) to obtain:
\begin{equation}
\mathcal{G}_{\Lambda\,t\tau}=-i\mu_{\Lambda\Sigma}\sqrt{-g}\mathcal{F}^{\Sigma\,\theta\phi}-\nu_{\Lambda\Sigma}\mathcal{F}^{\Sigma}_{t\tau}\
.
\end{equation}
Substituting our potentials and with some rearranging, this becomes:
\begin{equation}
\sqrt{-g}\mathcal{F}^{\Sigma\,\theta\phi}=i(\mu^{-1}\nu)^{\Sigma}_{\
\Gamma}\partial_{\tau}\psi^{\Gamma}
+i(\mu^{-1})^{\Sigma\Lambda}\partial_{\tau}\chi_{\Lambda}\ .
\label{fthetaphi}
\end{equation}
Since all other components of $\mathcal{F}$ are zero, we can write
the vector part of the Lagrangian as:
\begin{eqnarray}
2\sqrt{-g}\Bigg(\mu_{\Lambda\Sigma}\mathcal{F}^{\Lambda}_{t\tau}\mathcal{F}^{\Sigma}_{t\tau}g^{tt}g^{\tau\tau}
+\mu_{\Lambda\Sigma}\left(\sqrt{-g}\mathcal{F}^{\Lambda\,\theta\phi}\right)\left(\sqrt{-g}\mathcal{F}^{\Sigma\,\theta\phi}\right)
\frac{g_{\theta\theta}g_{\phi\phi}}{\left(\sqrt{-g}\right)^2}\nonumber\\
+\nu_{\Lambda\Sigma}\mathcal{F}^{\Lambda}_{t\tau}\left(\sqrt{-g}\mathcal{F}^{\Sigma\,\theta\phi}\right)g^{tt}g^{\tau\tau}
+\nu_{\Lambda\Sigma}\left(\sqrt{-g}\mathcal{F}^{\Lambda\,\theta\phi}\right)\mathcal{F}^{\Sigma}_{t\tau}g^{tt}g^{\tau\tau}\Bigg)\
.
\end{eqnarray}
Our metric ansatz gives:
\begin{eqnarray}
g^{tt}g^{\tau\tau}&=&-\frac{\sinh^4(c\tau)}{c^4}\nonumber\\
\frac{g_{\theta\theta}g_{\phi\phi}}{\left(\sqrt{-g}\right)^2}&=&\frac{\sinh^4(c\tau)}{c^4}\nonumber\\
\sqrt{-g}&=&\frac{c^4e^{-2U}\sin{\theta}}{\sinh^4(c\tau)}\ .
\end{eqnarray}
Thus we obtain:
\begin{eqnarray}
-2e^{-2U}\sin(\theta)\left(\mu_{\Lambda\Sigma}\mathcal{F}^{\Lambda}_{t\tau}\mathcal{F}^{\Sigma}_{t\tau}
-\mu_{\Lambda\Sigma}\left(\sqrt{-g}\mathcal{F}^{\Lambda\,\theta\phi}\right)\left(\sqrt{-g}\mathcal{F}^{\Sigma\,\theta\phi}\right)
\right.\nonumber\\
\left.+\nu_{\Lambda\Sigma}\mathcal{F}^{\Lambda}_{t\tau}\left(\sqrt{-g}\mathcal{F}^{\Sigma\,\theta\phi}\right)
+\nu_{\Lambda\Sigma}\left(\sqrt{-g}\mathcal{F}^{\Lambda\,\theta\phi}\right)\mathcal{F}^{\Sigma}_{t\tau}\right)\
. \label{vectorlangrangian2}
\end{eqnarray}
Since it is now clear that the $\theta$-dependance of the Lagrangian
factors out (as we would expect), we shall drop this leading factor
from now on. Our next task is to express the parenthetical term in
(\ref{vectorlangrangian2}) in terms of the electric and magnetic
potentials, using (\ref{empot}) and (\ref{fthetaphi}). Making the
appropriate substitutions, using the symmetry of $\mu$ and $\nu$ and
taking the real part of the action, the unquantized vector
Lagrangian is:
\begin{equation}
-2e^{-2U}\left(\partial_{\tau}\psi^{\Lambda}\ \ \
\partial_{\tau}\chi_{\Lambda}\right)
\left(\begin{array}{cc} (\mu+\nu\mu^{-1}\nu)_{\Lambda\Sigma} &\ \
(\nu\mu^{-1})^{\Sigma}_{\ \Lambda}\\
(\mu^{-1}\nu)^{\Lambda}_{\ \Sigma} &\ \ (\mu^{-1})^{\Lambda\Sigma}\\
\end{array}\right)
\left(\begin{array}{c}\partial_{\tau}\psi^{\Sigma}\\
\partial_{\tau}\chi_{\Sigma}\end{array}\right)\
.\label{vlmatrixform}
\end{equation}
The quantize this, we use the fact that it is manifestly independent
of $\psi$ and $\chi$, only their derivatives appear. Thus:
\begin{eqnarray}
\frac{\partial{\mathcal{L}}}{\partial\psi^{\Lambda}}=0&\Rightarrow&\frac{\partial{\mathcal{L}}}{\partial(\partial_{\tau}\psi^{\Lambda})}=q_{\Lambda}\nonumber\\
\frac{\partial{\mathcal{L}}}{\partial\chi_{\Lambda}}=0&\Rightarrow&\frac{\partial{\mathcal{L}}}{\partial(\partial_{\tau}\chi_{\Lambda})}=p^{\Lambda}\
.
\end{eqnarray}
In more condensed notation (\ref{vlmatrixform}) becomes:
\begin{equation}
-e^{-2U}(\partial_{\tau}\hat{\chi}_{\Lambda})(M^{-1})^{\Lambda\Sigma}(\partial_{\tau}\hat{\chi}_{\Sigma})\
. \label{quantcond}
\end{equation}
Here $(M^{-1})^{\Lambda\Sigma}(z,\bar{z})$ is the middle matrix in
(\ref{vlmatrixform}) and $\hat{\chi}_{\Lambda}$ is a vector whose
entries contain both $\chi_{\Lambda}$ and $\psi^{\Lambda}$, ordered
as in (\ref{vlmatrixform}). We can then use the constraints in
(\ref{quantcond}) to get:
\begin{equation}
\hat{p}^{\Lambda}=2e^{-2U}(M^{-1})^{\Lambda\Sigma}(\partial_{\tau}\hat{\chi}_{\Sigma})\
.
\end{equation}
$\hat{p}^{\Lambda}$ is a charge vector containing both the
$p^{\Lambda}$ and $q_{\Lambda}$ defined above. This implies that:
\begin{equation}
(\partial_{\tau}\hat{\chi}_{\Sigma})=\frac{1}{2}e^{2U}\hat{p}^{\Lambda}M_{\Lambda\Sigma}\
.
\end{equation}
Making this substitution in (\ref{vlmatrixform})\footnote{Note that
we now use the inverse of the previous matrix, and invert the order
of the electric and magnetic charges.} and combining all the parts
together gives:
\begin{equation}
\mathcal{L}=\left(\frac{\partial U}{\partial\tau}\right)^2
+G_{a\bar{a}}\partial_{\tau}z^a\partial_{\tau}\bar{z}^{\bar a}
+\frac{1}{4}\,e^{2U}\left(p^{\Lambda}\ \ q_{\Lambda}\right)
\left(\begin{array}{cc} (\mu+\nu\mu^{-1}\nu)_{\Lambda\Sigma} &\ \
-(\nu\mu^{-1})^{\Sigma}_{\ \Lambda}\\
-(\mu^{-1}\nu)^{\Lambda}_{\ \Sigma} &\ \ (\mu^{-1})^{\Lambda\Sigma}\\
\end{array}\right)
\left(\begin{array}{c}p^{\Sigma}\\ q_{\Sigma}\end{array}\right)\ .
\end{equation}
This is of the form (\ref{lagr}), with:
\begin{equation}
V_{BH}(z,\bar{z},p,q)=\frac{1}{4}\left(p^{\Lambda}\ \
q_{\Lambda}\right)
\left(\begin{array}{cc}(\mu+\nu\mu^{-1}\nu)_{\Lambda\Sigma} &\ \
-(\nu\mu^{-1})^{\Sigma}_{\ \Lambda}\\
-(\mu^{-1}\nu)^{\Lambda}_{\ \Sigma} &\ \
(\mu^{-1})^{\Lambda\Sigma}\\ \end{array}\right)
\left(\begin{array}{c}p^{\Sigma}\\ q_{\Sigma}\end{array}\right)\ .
\end{equation}

\section{Covariant Derivatives}
In \cite{Tripathy:2005qp}, the equation which is obtained from the
minimization of the effective potential of black hole is expressed
in the following way:
\begin{equation}\label{z1}
\partial_{a}V_{eff}=e^{K}\Big(G^{b\bar{c}}(\nabla_{a}\nabla_{b}W)\bar{\nabla}_{\bar{c}}\overline{W}
+2(\nabla_{a}W)\overline{W}+ (\partial_{a}G^{b\bar{c}})(\nabla_{b}W)
(\bar{\nabla}_{\bar{c}}\overline{W})\Big)=0\ .
\end{equation}
$\nabla_{a}W=(\partial_{a}+(\partial_{a}K))W$, and
also\footnote{Note that $\nabla_{b}W={\mathcal{D}}_{b}W$. For the
scalars which are defined on the moduli space, ${\mathcal{D}}$ and
$\nabla$ act in the same way.}:
\begin{equation}\label{z2}
\nabla_{a}\nabla_{b}W=(\partial_{a}+\partial_{a}K)\nabla_{b}W\ .
\end{equation}
Now, consider the full covariant derivative acting on
$\mathcal{D}_bW$:
\begin{equation}\label{z3}
{\mathcal{D}}_{a}{\mathcal{D}}_{b}W=\Big((\partial_{a}+
\partial_{a}K)\delta^{d}_{b}-\Gamma^{d}_{ab}\Big)
{\mathcal{D}}_{d}W\ ,
\end{equation}
$\Gamma^{d}_{ab}$ are the Christoffel symbols of Levi-Civita
connection on the moduli space. We can thus rewrite the last term of
(\ref{z1}) in the following way:
\begin{eqnarray}\label{z4}
(\partial_{a}G^{b\bar{c}}){\mathcal{D}}_{b}W&=&
(\partial_{a}G^{b\bar{c}})\delta^{d}_{b}{\mathcal{D}}_{d}W=
(\partial_{a}G^{b\bar{c}})(G_{b\bar{e}}G^{\bar{e}d}){\mathcal{D}}_{d}W\nonumber\\
&=&-G^{b\bar{c}}(\partial_{a}G_{b\bar{e}})G^{\bar{e}d}{\mathcal{D}}_{d}W=-G^{b\bar{c}}\Gamma^{d}_{ab}{\mathcal{D}}_{d}W\
,
\end{eqnarray}
where we used the expression
$\Gamma^{d}_{ab}=G^{d\bar{e}}\partial_{a}G_{\bar{e}b}$ for
Christoffel symbols of a metric compatible K\"{a}hler manifold.
Using this (\ref{z1}) becomes:
\begin{eqnarray}\label{z5}
\partial_{a}V_{eff}&=&e^{K}\Big(G^{b\bar{c}}\big[(\partial_{a}+\partial_{a}K)\delta^{d}_{b}-
\Gamma^{d}_{ab}\big]{\mathcal{D}}_{d}W\bar{{\mathcal{D}}}_{\bar{c}}\overline{W}
+2({\mathcal{D}}_{a}W)\overline{W}\Big)\nonumber\\
&=&e^{K}\Big(G^{b\bar{c}}({\mathcal{D}}_{a}{\mathcal{D}}_{b}W)\bar{D}_{\bar{c}}\overline{W}+
2({\mathcal{D}}_{a}W)\overline{W}\Big)\ .
\end{eqnarray}
It is clear that the above equation has the appropriate covariant
form.

\section{Hodge-decomposition of 3-form flux}
As we discussed in Section 1, (\ref{e5}) (or equivalently
(\ref{e4})) is the Hodge-decomposition of the flux form and this
equation by itself does not contain any information about the moduli
at horizon of the black hole. In other words, (\ref{e5}) does not
pick any point of the moduli space up and it is trivially satisfied.
Of course, at first glance, it might be thought that this equation
can be solved for the moduli; after all the r.h.s. of (\ref{e5}) is
a function of moduli $z^{a}$ and the l.h.s. only includes a set of
integers. Nevertheless, the r.h.s. of (\ref{e5}) is always
independent of the moduli coordinates. We can explicitly show that
this is the case for the specific example considered in section 5.1.

Since we have already computed all the ingredients of (\ref{e5}) in
section 5.1, we can easily calculate the r.h.s. Considering
(\ref{a9}), (\ref{a13}), and (\ref{c9}), the second term of the
r.h.s. of (\ref{e5}) is given by:
\begin{eqnarray}\label{g1}
G^{\bar{a}b}\overline{{\mathcal{D}}}_{\bar{a}}\overline{Z}{\mathcal{D}}_{b}\Pi=-\frac{i}{6}\Big(
M^{ab}+\frac{12}{M}p^{a}p^{b}t^{2}\Big)\Big(+3itD_{a}(2-\frac{W}{2Dt^{2}})\Big)
\left(\begin{array}{c}
\frac{3}{2}i\frac{D_d}{D}\frac{1}{t} \\
\delta^{c}_{d}-\frac{3}{2}p^{c}\frac{D_{d}}{D} \\
\frac{3}{2}D_{d}t^{2} \\
6iD_{cd}t-\frac{9}{2}i\frac{D_{d}D_{c}}{D}t \\
\end{array} \right)\ .
\end{eqnarray}
After simplifying the above expression, we obtain for the second
term of the r.h.s. of (\ref{e5}):
\begin{eqnarray}\label{g2}
2\mbox{Im}\Big(G^{\bar{a}b}\overline{{\mathcal{D}}}_{\bar{a}}\overline{Z}{\mathcal{D}}_{b}\Pi\Big)=
-\Big(1-\frac{W}{4Dt^{2}}\Big) \left(
\begin{array}{c}
0 \\
-p^{c}\\
3Dt^{2}\\
0\\
\end{array}
\right)\ .
\end{eqnarray}
Then, from (\ref{a7}), we find:
\begin{eqnarray}\label{g3}
2\mbox{Im}\Big[\
Z\overline{\Pi}+G^{\bar{a}b}\overline{{\mathcal{D}}}_{\bar{a}}\overline{Z}{\mathcal{D}}_{b}\Pi\
\Big]=\left(
        \begin{array}{c}
          0\\
          p^{c} \\
          W-3Dt^{2}\\
          0 \\
        \end{array}
      \right)\ ,
\end{eqnarray}
where $W-3Dt^{2}$ is nothing but $q_{0}$. Therefore, the final
answer is precisely the set of electric and magnetic charges $h$.
Hence (\ref{e5}) is true for any arbitrary point of the moduli space
and this equation does not pick up any specific point of the moduli
space.

\section{More Details on Mirror Quintic}
\setcounter{equation}{0} In this appendix, for completeness, we
present the formulas used in Section 5.3. In the vicinity of
$\psi=0$, the period vector in the Picard-Fuchs basis is obtained by
solving the Picard-Fuchs equation (\cite{Candelas:1990rm}). The
fundamental period vector is given by the following series:
\begin{eqnarray}\label{ap1}
\omega_{0}(\psi)=-\frac{1}{5}\sum_{m=1}^{\infty}\frac{\alpha^{2m}\Gamma(m/5)(5\psi)^{m}}{\Gamma(m)\Gamma^{4}(1-m/5)}\
,
\end{eqnarray}
where $\alpha=e^{2\pi i/5}$. By changing the basis, we can obtain
the period vector in symplectic form. It turns out that the period
vector in the symplectic basis is given by:
\begin{eqnarray}\label{ap2}
\Pi=\tilde{m}\bar{\omega}(\psi)\ .
\end{eqnarray}
$\bar{\omega}$ itself is given by:
\begin{eqnarray}\label{ap3}
\bar{\omega}=-\frac{1}{\psi}\Big(\frac{2\pi i}{5}\Big)^{3} \left(
  \begin{array}{c}
    \omega_{2} \\
    \omega_{1} \\
    \omega_{0} \\
    \omega_{4} \\
  \end{array}
\right)\ .
\end{eqnarray}
Here $\omega_{i}=\omega_{0}(\alpha^{i}\psi)$ and matrix $\tilde{m}$
is:
\begin{eqnarray}\label{ap4}
\tilde{m}=\left(
\begin{array}{cccc}
-1 & 0 & 8 & 3 \\
0 & 1 & -1 & 0 \\
-\frac{3}{5} & -\frac{1}{5} & \frac{21}{5} & \frac{8}{5} \\
0 & 0 & -1 & 0 \\
\end{array}
\right)\ .
\end{eqnarray}
It is convenient to define the coefficients $c_{m}$ and column
vectors $p_{m}$ as:
\begin{eqnarray}\label{ap5}
c_{m}=\frac{\Gamma((m+1)/5)5^{m+1}}{\Gamma(m+1)\Gamma^{4}((4-m)/5)}\
\ ,\ \ p_{m}= \left(
  \begin{array}{c}
    \alpha^{4(m+1)} \\
    \alpha^{3(m+1)} \\
    \alpha^{2(m+1)} \\
    \alpha^{(m+1)} \\
  \end{array}
\right)\ .
\end{eqnarray}
Recalling (\ref{ff6}), in which $\Sigma$ is
\begin{eqnarray}\label{ap6}
\Sigma=\left(
         \begin{array}{cccc}
           0 & 0 & 1 & 0 \\
           0 & 0 & 0 & 1 \\
           0 & -1 & 0 & 0 \\
           -1 & 0 & 0 & 0 \\
         \end{array}
       \right)\ ,
\end{eqnarray}
we can find the K\"{a}hler potential as a power series in terms of
$\psi$:
\begin{eqnarray}\label{ap7}
K=C_{0}-\log\Big(1+(2-\sqrt{5})\frac{c_{1}^{2}}{c_{0}^{2}}|\psi|^{2}-
(2-\sqrt{5})\frac{c_{2}^{2}}{c_{1}^{2}}|\psi|^{4}+\cdots\Big)\ .
\end{eqnarray}
$C_{0}$ is the following additive constant:
\begin{eqnarray}\label{ap8}
C_{0}=-\log\Big[\sqrt{2+2\sqrt{5}}\Big(\frac{c_{0}}{5}\Big)^{2}\Big(\frac{2\pi}{5}\Big)^{6}\Big]\
.
\end{eqnarray}
The K\"{a}hler metric and its inverse are then expressed in terms of
$\psi$ as:
\begin{eqnarray}\label{ap9}
&&G_{\psi\bar{\psi}}=-(2-\sqrt{5})\frac{c_{1}^{2}}{c_{0}^{2}}+2(2-\sqrt{5})\Bigg[\Big(\frac{c_{1}}{c_{0}}\Big)^{4}
+2\Big( \frac{c_{2}}{c_{0}}\Big)^{2}|\psi|^{2}\Bigg]+\cdots\
,\\
&&G^{\psi\bar{\psi}}=-\frac{c_{0}^{2}}{c_{1}^{2}(2-\sqrt{5})}\Bigg[1+\Big(2(2-\sqrt{5})
\frac{c_{1}^{2}}{c_{0}^{2}}+4\frac{c_{2}^{2}}{c_{1}^{2}}\Big)|\psi|^{2}+\cdots\Bigg]\
.
\end{eqnarray}

\end{document}